\documentclass{elsart5p}

\usepackage{amsfonts}
\usepackage{amsmath}
\usepackage{amssymb}
\usepackage{graphicx}

\graphicspath{{graphics/}}

\theoremstyle{plain}
\newtheorem{thrm}{Theorem}[section]
\newtheorem{lema}{Lemma}[section]

\theoremstyle{definition}
\newtheorem{defin}{Definition}[section]
\newtheorem{exmpl}{Example}[section]

\newcommand{\eat}[1]{}

\begin{document}

\begin{frontmatter}



\title{A procedural framework and mathematical analysis for solid sweeps}


\author{Bharat Adsul, Jinesh Machchhar, Milind Sohoni}

\address{Department of Computer Science and Engineering, Indian Institute of Technology Bombay, Mumbai-400076, India}

\begin{abstract}
Sweeping is a powerful and versatile method of designing objects.
Boundary of volumes (henceforth envelope) obtained by sweeping solids have
been extensively investigated in the past, though, obtaining an accurate
parametrization of the envelope remained computationally
hard. The present work reports our approach to this problem as
well as the important problem of identifying self-intersections within
the envelope. Parametrization of the envelope is, of course, necessary for
its use in most current CAD systems. We take the more interesting case
when the solid is composed of several faces meeting smoothly. We show that
the face structure of the envelope mimics locally that of the solid.
We adopt the procedural approach at defining the geometry
in this work which has the advantage of being accurate as
well as computationally efficient. The problem of
detecting local self-intersections is central to a robust implementation
of the solid sweep. This has been addressed by computing a
subtle mathematical invariant which detects self-intersections, and which
is computationally benign and requires only point queries.
\end{abstract}

\end{frontmatter}



\section{Introduction} \label{introductionSec}

In this paper we focus on the problem of computing an accurate parametrization of the boundary of the volume obtained by sweeping a solid in $\mathbb{R}^3$ along a trajectory and that of detecting local self-intersections in the envelope.  Sweeping is an operation of fundamental importance in geometric design. It has applications like numerically controlled machining verification \cite{sede, sede2, completeSweep} and robot motion planning \cite{workspace, workspace2}.  There have been several approaches to computing the boundary of swept volumes in the past. The works \cite{jacobian, geoModeling} formulate the problem using the rank deficiency of the Jacobian, \cite{sede, sede2, sede3} compute the envelope by solving sweep differential equations,  \cite{classifyPoints} uses inverse-trajectories for deriving a point membership test for a point to belong to the envelope.  In \cite{screwMotion} the authors give a close approximation of the envelope by restricting the trajectories to piecewise screw motions.  

Despite the extensive research done in the past in this area, computing an accurate parametrization of the envelope has remained an unsolved problem due to known mathematical and computational difficulties \cite{voidDetection}.  In this work we attempt to arrive at an accurate parametrization of the envelope through the procedural approach, which is an abstract way of defining surfaces and curves when closed form formulae are not available.  The procedural paradigm exploits the fact that from the users point of view, a parametric surface is just a map from $\mathbb{R}^2$ to $\mathbb{R}^3$ and hence can be represented in a computer by a procedure which takes as input $(u,v) \in \mathbb{R}^2$ and returns $(x,y,z) \in \mathbb{R}^3$.  Higher order derivatives of the surface can be returned similarly.  The definition of splines through the De Casteljau's algorithm is an example of procedural parametrization.  The authors in \cite{procedural} compute the intersection curve of two parametric surfaces by procedural approach.

The second problem that we tackle in this paper is that of detecting local self-intersections. Self-intersections cause anomalies in the envelope.  There have been rather few attempts at solving this problem in the past.  The paper \cite{trimming} proposes an efficient and robust method of detecting global and local self-intersections by checking whether the inverse-trajectory of a point intersects the solid. In the paper \cite{selfIntersections} global and local self-intersections are detected by computing intersection of curves of contact at discrete time steps.  This has the disadvantage of being computationally expensive.  In the paper \cite{completeSweep2} self-intersections are accurately quantified and detected but their method is limited to sweeping tools for NC machining verification.  The method employed by \cite{peternell} for detecting local self-intersections is based on point set data and could be computationally expensive. In \cite{errorBounds} the author solves the problem of detecting local self-intersections for sweeping planar profiles.  In this paper we propose a novel test for detecting local self-intersections which is based on a subtle mathematical invariant of the envelope.  It has the advantage of being computationally efficient and requires only point queries.

The paper is organized as follows.  In Section~\ref{definitionSec} we describe the input to the sweeping algorithm, in Section~\ref{architecture} we discuss the overall framework for the computation of the envelope, in Sections~\ref{lsiSec} and \ref{lsiSec2} we study the mathematical structure of the envelope and quantify local self-intersections, giving a test for detecting them, in Section~\ref{smoothSec} we analyse the case when the envelope is free from self-intersections.  In Section~\ref{computationSec} we describe the algorithm for computing the procedural parametrization of the envelope.  We conclude the paper in Section~\ref{conclusion}.


\section{Preliminaries} \label{definitionSec}

This section outlines the basic representational structures associated with 
the  problem.
Subsection~\ref{brepSubSec} describes the boundary representation of a solid
which is typical to many CAD systems and subsequent sections, the    
basic inputs and outputs of the sweep algorithm. 
In Subsection~\ref{trajSubSec} we define the trajectory 
in $\mathbb{R}^3$ along which the solid is swept. Next, in 
subsection~\ref{envlSubSec} we define the various mathematical sub-entities 
which make up the envelope. 

\subsection{Boundary representation of a solid} \label{brepSubSec}
Boundary representation, also known as Brep, is a popular and standard method of representing a `closed' solid $M$ by its boundary $\partial M$.  The boundary $\partial M$ separates the interior of 
$M$ from the exterior of $M$.  $\partial M$ is represented using a set of \emph{faces}, \emph{edges} and \emph{vertices}.  See figure~\ref{brepFig} for a Brep of
a solid where different faces are coloured differently.  Faces meet in edges and edges meet in vertices.  The Brep of a solid consists of two interconnected pieces of information, viz. 
the geometric and the topological. 

\noindent
{\bf Geometric information:}
This consists of geometric entities, namely, vertices, edges and faces. A vertex is simply a point in $\mathbb{R}^3$. An edge is obtained by restricting the underlying parametric curve by a pair of vertices.  A parametric curve in $\mathbb{R}^3$ is a continuous map $\gamma: \mathbb{R} \to \mathbb{R}^3$. The curve $\gamma$ is called regular at $s_0 \in \mathbb{R}$ if $\gamma$ is differentiable and $\frac{d \gamma}{ds}|_{s_0} \neq \bar{0}$. Here $s$ is the parameter of the curve. An edge is derived from the underlying curve by suitably restricting the parameter 
$s$ to 
an interval $[a,b]$.  Further, it is required that the edge (more precisely, the underlying curve) is regular at all points in the interval $[a,b]$ and devoid of
self-intersections.

Similarly, a face is obtained by restricting the underlying parametric surface by a set of edges.
A parametric surface is a continuous map $S:\mathbb{R}^2 \to \mathbb{R}^3$.  The surface $S$ is said to be regular at $(u_0,v_0) \in \mathbb{R}^2$  if $S$ is differentiable 
and $\frac{\partial S}{\partial u}|_{(u_0,v_0)} \in \mathbb{R}^3$ and $\frac{\partial S}{\partial v}|_{(u_0,v_0)} \in \mathbb{R}^3$ are linearly independent.
Here $u$ and $v$ are the parameters of $S$. A
face is derived from a surface by suitably restricting the parameters $u$ and $v$ inside a `domain'. As expected, it is required that the face is regular
at all points in the domain and devoid of self-intersection.

\noindent
{\bf Topological information:}
The topological/combinatorial information consists of spatial relationships \eat{(like adjacency, orientation)} between different geometric entities, i.e., the 
adjacency between faces, the incidence relationships between faces and edges and so on. In figure~\ref{brepFig}, for example, the orange and the green face are adjacent.
Another important component is the orientation for each face, that is, a consistent choice of outward-normal for that face.
The orientation of a regular face is a choice of a unit normal from amongst $\frac{S_u \times S_v}{\| S_u \times S_v \|}$ and $-\frac{S_u \times S_v}{\| S_u \times S_v \|}$ 
where $S$ is the underlying parametric surface.
All the faces bounding the solid are oriented so that the unit normal at each point on each face is pointing towards the exterior of the solid.

Conceptually, a Brep through its `global' topological information glues the `local' geometric entities which come equipped with associated 
mathematical parametrizations. Note that, the regularity assumptions on the geometric entities guarantee that the tangent space at every point on an edge or a face is of the 
right dimension. Typically, one also imposes higher-order `parametric' continuity requirements which are denoted by $C^k$ where $k$ refers to the order of continuity.
For the sake of simplicity, throughout this paper, we will assume that the edges and faces bounding the solid are regular of class $C^k$ for some $k \geq 2$, 
i.e. the underlying parametrizations are twice differentiable with continuous second order derivatives.
Note that, however, these do not rule out, e.g.,  adjacent faces meeting along sharp edges. 
\eat{Absence of such sharp features are modeled by so called `geometric' 
continuity requirements and are denoted by $G^k$. (reference?) Note that, in the solid in figure.. , all the faces satisfy $G^1$ continuity. This means
that, even across faces, the normal varies continuously.}

\begin{figure}
 \centering
 \includegraphics[scale=0.52]{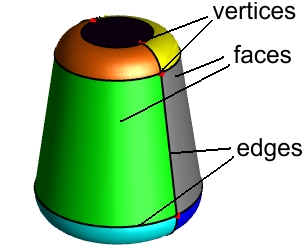}
 \caption{A Brep of a solid}
 \label{brepFig}
\end{figure}

\eat{onwe do not require adjacent faces to meet smoothly along the common edge. 

\eat{adjacency between faces and edges 
between a face and an edge is stored in a data-structure called \emph{coedge}.  Fig.~\ref{brepFig} shows a Brep of a solid.}

All the faces are assigned a consistent orientation.  

\begin{figure}
 \centering
 \includegraphics[scale=1.0]{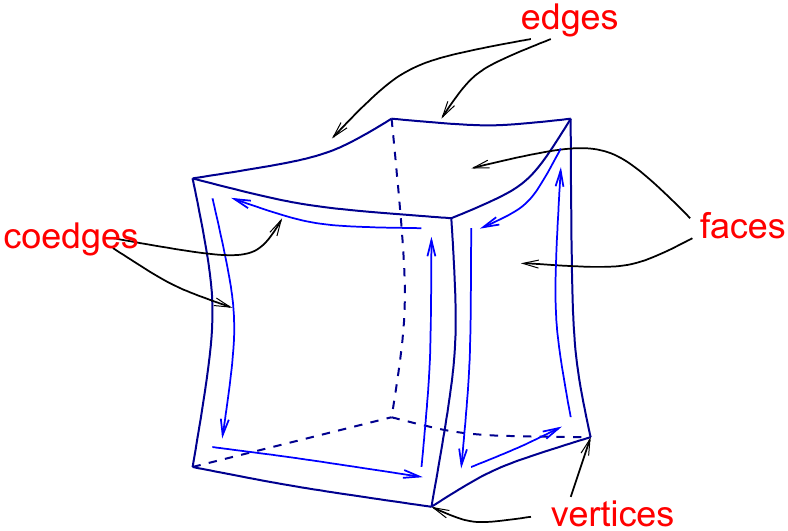}
 \caption{A Brep of a solid}
 \label{brepFig}
\end{figure}

The tangent space at every point of a regular surface is two dimensional.  If the tangent space at a point on a curve or surface is rank deficient, such a point is said to be a singular point.  We assume that the boundary of the solid is free from self-intersections.  In summary, for the sake of mathematical precision, we may assume that the solid is a compact 3-dimensional manifold with boundary while its boundary is a compact 2-dimensional manifold.  If $X$ is a $k$-dimensional differentiable manifold, we will denote the tangent space at a point $p \in X$ by $\mathcal{T}_{X}(p)$.}

\subsection{A trajectory in $\mathbb{R}^3$}	\label{trajSubSec}
A trajectory in $\mathbb{R}^3$ is a 1-parameter family of rigid motions in $\mathbb{R}^3$ defined as follows.
\begin{defin} \label{trajectoryDef}
A \emph{trajectory} in $\mathbb{R}^3$ is specified by a map $h:[0,1] \rightarrow (SO(3), \mathbb{R}^3),  h(t) = (A(t), b(t))$ where  $ A(t) \in SO(3) \footnote{$SO(3)=\{X \mbox{ is a 3 $\times$3 real matrix} |X^tX = I, det(X)=1  \}$ is the special orthogonal group, i.e. the group of rotational transforms.}, b(t) \in \mathbb{R}^3, A(0) = I, b(0) = 0$.    The parameter $t$ in this definition represents time.    
\end{defin}
For technical convenience, we assume that $h$ is of class $C^k$, for some $k \geq 2$.  

\subsection{Boundary of the swept volume}	\label{envlSubSec}

We begin by giving an intuitive description of the boundary of the swept volume.  We will formalize these notions in Section~\ref{lsiSec}.  Let $M$ be a solid being swept along a given trajectory $h$.  By abuse of notation, a point in $M$ will  mean a point in the interior of $M$ or on the boundary $\partial M$ of $M$.  We denote by $M_t$ the position of $M$ at time $t \in [0, 1]$, i.e. $M_t = \{ A(t)x + b(t) | x \in M\}$, and by $\partial M_t$, the boundary of $M_t$. Then $\displaystyle \bigcup_{t \in [0,1]} M_t$ is the volume swept by $M$ during this operation.  Our goal is to compute the boundary of this swept volume as a Brep, which we will refer to as the \emph{envelope}.  For a fixed point $x \in M$, consider the trajectory of $x$ as the map $y:[0,1] \to \mathbb{R}^3$  given by $y(t)= A(t)x + b(t)$. The trajectory of $x$ describes the motion $x$ in $\mathbb{R}^3$ under the given trajectory $h$. 
Clearly, if $x$ is in the interior of $M$, no point in the image of the trajectory of $x$ can be on the envelope. Further, at a particular time instant $t_0$, only a 
subset of points on $\partial M_{t_0}$ will lie on the envelope.  The union of such points for all $t_0 \in [0,1]$ gives the final envelope.  

It is clear that, at a given time instant $t_0$, only a part of $\partial M_{t_0}$ is in `contact' with the envelope. To make this more clear, fix a point 
$x \in \partial M$ and the trajectory $y$ of $x$. The derivative of the trajectory of $x$ at a given time instant $t_0$, that is, 
$\frac{dy}{dt}|_{t_0}$ gives the velocity of $x$ at $t_0$. It is easy to show that (cf, Section~\ref{lsiSec}) $x$ (more precisely, $y(t_0)$) is in contact with the envelope at time $t_0$ only if 
the velocity of $x$ at $t_0$ is in the `tangent-space' of $\partial M_{t_0}$ at $y(t_0)$. In a generic situation, the set of points of $\partial M_{t_0}$ 
which are in contact with the envelope will be the curve-of-contact. 
The union of these curves-of-contact is called the 
\emph{contact-set} or the \emph{running envelope}.  
Clearly, the total envelope and the contact-set are closely related. If all
goes well, the 
envelope is obtained from the contact-set by `capping' it by appropriate
parts of $M_0 $ and $M_1 $, the object at times $t=0$ and $t=1$. 
But all may not go well. The detection of anomalies is central to the use
of the algorithm in industrial situations and is an important objective of 
this paper.

\begin{figure*}
 \centering
 \includegraphics[scale=0.75]{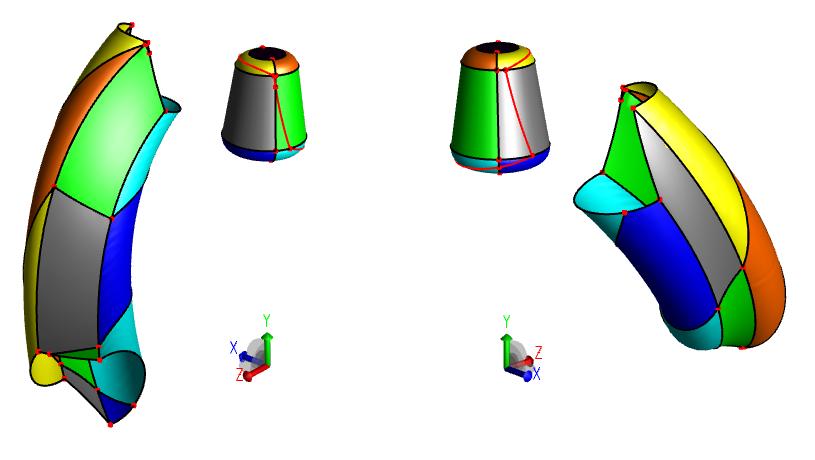}
 \caption{A solid is swept along a helical trajectory}
 \label{sweepExample}
\end{figure*}

\section{Our Approach/Framework} \label{architecture}
In this section we briefly describe the overall framework for computing the sweep surface i.e. the envelope as a Brep. We continue using the notation from
the previous section where $M$ denotes the Brep/solid and $h$ denotes the trajectory along which $M$ is swept.  

The naive approach to computing the envelope would be to discretize time, i.e, 
to construct a sequence $T=\{ 0=t_1 ,\ldots ,t_k =1 \} $ and constructing the 
{\em approximate} envelope as $E'=\cup_i M_{t_i}$, the union of the translates.
The next step would be to construct a {\em smooth} version $E''$ 
 of $E'$ above, by 
some fitting operation. However, this approach has several issues--(i) 
computation of $E'$ leads to unstable booleans of two very close-by objects, leading to sliver-faces, and (ii) the fit of $E''$ to the actual $E$ depends on
a dense enough choice of $T$ which compounds problem (i) above. There are other
options, but problems remain. 

In this work, we propose a novel approach based on the {\em procedural 
paradigm} (cf~\cite{sohoni, procedural}) which has gained ascendance in 
many numerical kernels, e.g., ACIS (cf~\cite{acis}). 

We now describe the basic architecture for our algorithm. For this, we 
use a running example referred to in figure~\ref{brepFig} and figure~\ref{sweepExample}. The object to be swept is 
$M$ as in figure~\ref{brepFig}, and the output contact-set is $\mathcal{C}$ as shown in figure~\ref{sweepExample}. The trajectory is 
roughly helical with a compounded rotation. 
 
\begin{enumerate}
\item {\it A natural correspondence between the entities of $M$ and the entities of $\mathcal{C}$.}

Every point $p$ of the envelope comes from a curve of contact on $M_t $, for 
some $t$, and therefore belongs to some entity of $M$, i.e., a vertex, edge
or face. This sets up the correspondence between entities of $M$ and 
those of $\mathcal{C}$. The procedural approach attaches a common evaluation method 
to each such entity. Fig2 illustrates this correspondence.   
Faces of $\mathcal{C}$ which are generated by 
a particular face of $M$ are shown in same colour.  Curve-of-contact at time $t=0$ is shown imprinted on the solid in red.

Along with the geometric definition of each entity, we must also construct 
the topological data to go with it. This data is constructed by observing
that there is a local homeomorphism between a point on $\mathcal{C}$ and a suitable
point on $M$. 

\item {\it Accurate parametrizations of the geometric entities of $\mathcal{C}$ with `time' as one of the central parameters.}

This is achieved through the procedural paradigm in which all key attributes/features of the geometric entities are made available through a set
of associated procedures (cf~\cite{sohoni, procedural}). In our case, these procedures are based on Newton-Raphson solvers. This is the focus of Section~\ref{computationSec}.

\eat{*** Describe at high-level how the underlying parametrizations are obtained *** }

\item {\it Topological and regularity analysis of $\mathcal{C}$.}

It is quite common to have a sweeping operation in which the resulting envelope/contact-set $\mathcal{C}$ self-intersects. These self-intersections can
be broadly classified into global and local self-intersections (see figure~\ref{globalLocalFig}).  Once an accurate `local' parametrization (as in step 2) of the faces of $\mathcal{C}$ is obtained, 
in principle, global self-intersections can be detected and dealt with by well-known (cf \cite{procedural}) surface-surface intersection solvers.  A more subtle mathematical 
issue is that of detecting singularities and local self-intersections. This is addressed in Sections~\ref{lsiSec} and \ref{lsiSec2}.

\eat{but local self-intersections pose a challenge.  Another source of difficulty in constructing the Brep of the envelope is in obtaining a good parameterization of the edges and faces on the envelope.  In this paper we address two issues, namely, (1) detecting local self-intersections and (2) obtaining a parameterization of faces.  The rest of the computational aspects in computing the Brep of the envelope will be addressed in later work.}

\eat{The resulting envelope may have self-intersections, which can be broadly classified into global and local self-intersections.  In principle, global self-intersections can be detected by surface-surface intersection but local self-intersections pose a challenge.  Another source of difficulty in constructing the Brep of the envelope is in obtaining a good parameterization of the edges and faces on the envelope.  In this paper we address two issues, namely, (1) detecting local self-intersections and (2) obtaining a parameterization of faces.  The rest of the computational aspects in computing the Brep of the envelope will be addressed in later work.}

\eat{** Describe the self-intersection issues that may arise. also, highlight the difference between global vs local self-intersections.}

\end{enumerate}

We are now in a position to define the scope of this paper. In this work we describe in detail tasks (ii) and (iii) described above, namely, detecting local 
self-intersections and obtaining a procedural parameterization of faces.  
The focus of Section~\ref{smoothSec} is task (i) in the interesting case when $M$ is composed of faces meeting smoothly.
Other architectural aspects and solid-modelling implementation will be addressed in a later work.

\eat{** Once again summarize the contributions of our work and outline the scope of THIS paper vis-a-vis above three items.}

\eat{
In this section we describe the overall architecture of the computational framework for computing the Brep of the envelope obtained by sweeping the input solid $M$ along the input trajectory.  This means computing various geometric entities like vertices, edges and faces on the envelope as well as computing the topological information like adjacency between these geometric objects.  The resulting envelope may have self-intersections, which can be broadly classified into global and local self-intersections.  In principle, global self-intersections can be detected by surface-surface intersection but local self-intersections pose a challenge.  Another source of difficulty in constructing the Brep of the envelope is in obtaining a good parameterization of the edges and faces on the envelope.  In this paper we address two issues, namely, (1) detecting local self-intersections and (2) obtaining a parameterization of faces.  The rest of the computational aspects in computing the Brep of the envelope will be addressed in later work.}


\section{Mathematical structure of the contact-set} \label{lsiSec}

In this section we will study in detail the mathematical structure of the boundary of the volume obtained by
sweeping the solid $M$ along the trajectory. For simplicity, we work with a single parametric surface patch 
$S$ and analyse the sweep of $S$ under the trajectory $h$. As explained before,
we assume that both $S$ and $h$ are regular of class $C^k$ for $k \geq 2$, and are devoid of self-intersections. 
In section~\ref{smoothSec}, we lift the results of
this section to the interesting case when $M$ is composed of several faces/surfaces meeting smoothly.
For later use, we introduce the following notation: the tanget space to a manifold $X$ at a point $p \in X$ will be denoted by $\mathcal{T}_{X}(p)$.

We begin with the formal definition of the sweep map.

\begin{defin} \label{sweepDef}
Given $S$ and $h$, the \emph{sweep} is defined as a map $\sigma : \mathbb{R}^2 \times [0,1] \to \mathbb{R}^3$ given by $\sigma(u,v,t) = A(t)S(u,v) + b(t)$.
\end{defin}
Here $u,v$ are the parameters of $S$.  The position of a point $S(u_0,v_0)$ on surface $S$ at time $t_0$ will be given by $\sigma(u_0,v_0,t_0)=A(t_0)S(u_0,v_0)+b(t_0)$ and the velocity of the point $S(u_0,v_0)$ at time $t_0$ will be given by $V(u_0,v_0,t_0)=\frac{\partial \sigma}{\partial t} |_{(u_0,v_0,t_0)} = A'(t_0)S(u_0,v_0) + b'(t_0)$, where $'$ denotes derivative with respect to $t$.  If $N(u_0,v_0)$ is the unit (outward) normal to $S$ at $(u_0,v_0)$, then the unit normal to $S_{t_0}$ at $(u_0,v_0)$ is given by $\hat{N} = A(t_0)N(u_0,v_0)$ where, $S_{t_0} = \{ A(t_0)S(u,v) + b(t_0) | (u,v) \in \mathbb{R}^2 \}$ is the position  of the surface at time instant $t_0$.  
In order to formally define the contact-set, we look at the extended sweep in $\mathbb{R}^4$ in which the fourth dimension is time~\cite{trimming}.

\begin{defin} \label{extSweepDef}
Given $S$ and $h$, the \emph{extended sweep} is defined as a map $\hat{\sigma} : \mathbb{R}^2 \times [0,1] \to \mathbb{R}^4$ given by $\hat{\sigma}(u,v,t) = (\sigma(u,v,t), t)$.
\end{defin}

Thus, the sweep $\sigma$ is clearly the extended sweep $\hat{\sigma}$ composed with the projection map along the $t$-dimension.  Denoting partial derivatives using a subscript, we note that $\hat{\sigma}_u, \hat{\sigma}_v \mbox{ and } \hat{\sigma}_t$ are linearly independent for all $(u,v,t) \in \mathbb{R}^2 \times [0,1]$ and $\hat{\sigma}$ is injective. Hence the image of $\hat{\sigma}$ is a 3-dimensional manifold.  We now define the contact-set.
\eat{obtained by sweeping a given solid boundary $S$ along a given trajectory$h$.}

\begin{defin} \label{contactSetDef}
Given $S$ and $h$, the \emph{contact-set} is the set of points $\sigma(u_0,v_0,t_0)$ such that the line  $\{(x_0,y_0,z_0,t) \in \mathbb{R}^4| \sigma(u_0,v_0,t_0) = (x_0,y_0,z_0), t \in [0,1]  \}$ is tangent to (the image of) $\hat{\sigma}$ at $\hat{\sigma}(u_0,v_0,t_0) = (x_0, y_0, z_0, t_0)$.  We will denote the contact-set by $\mathcal{C}$.
\end{defin}

We will refer to the domain of the map $\sigma$ as the \emph{parameter space} and the co-domain as the \emph{object space}.  Consider now the following function $f: \mathbb{R}^2 \times [0,1] \to \mathbb{R}$ given by 
\begin{align} \label{envlEq}
f(u,v,t) = \left < V(u,v,t), \hat{N}(u,v,t) \right > 
\end{align}
Recalling that $V(u,v,t)$ is the velocity of the point $S(u,v)$ at time $t$ and $\hat{N}(u,v,t)$ is the normal to $S_t$ at $(u,v)$, we look at the zero-set of this function in the parameter space.

\begin{defin} \label{funnelDef}
The \emph{funnel} $\mathcal{F}$ is defined as the zero-set of the function $f$ specified in Eq.~\ref{envlEq}, i.e.,  $\mathcal{F} = \{(u,v,t) \in \mathbb{R}^2 \times [0,1] |f(u,v,t) = 0 \}$.
\end{defin}

In other words, if a point $p = (u_0,v_0,t_0) \in \mathcal{F}$, then the velocity at the point $\sigma(p)$ lies in the tangent space of $S_{t_0}$ at $\sigma(p)$.  The following lemma shows that the contact-set is precisely the image of the funnel through the sweep map.

\begin{lema} \label{envlLem}
$\sigma(\mathcal{F}) = \mathcal{C}$
\end{lema}
\emph{Proof.}  Skipped here.
\eat{
Suppose $p = (u_0, v_0, t_0)$, and $\sigma(p) \in \mathcal{C}$.  Then by definition~\ref{contactSetDef}, the line $l =  \{(x_0,y_0,z_0,t) \in \mathbb{R}^4| \sigma(u_0,v_0,t_0) = (x_0,y_0,z_0), t \in \mathbb{R}  \}$ is tangent to $\hat{\sigma}$ at $(u_0, v_0, t_0)$.  Hence the tangent space of $l$ is a subspace of the tangent space of $\hat{\sigma}$ at $(u_0,v_0,t_0)$. The tangent space of line $l$ is spanned by the vector $(0, 0, 0, 1)$ while the tangent space of $\hat{\sigma}$ is spanned by the set $\{\hat{\sigma}_u, \hat{\sigma}_v, \hat{\sigma}_t \}$ where, $\hat{\sigma}_u = (\sigma_u, 0)$, $\hat{\sigma}_v = (\sigma_v, 0)$ and $\hat{\sigma}_t = (\sigma_t, 1)$.  Hence, $(0, 0, 0, 1) = a\cdot (\sigma_u, 0) + b \cdot (\sigma_v, 0) + c \cdot (\sigma_t, 1)$ for $a,b,c \in \mathbb{R}, a,b,c$ not all zero.  This implies that $c=1$ and $\sigma_t = -a \cdot \sigma_u - b \cdot \sigma_v$.  Since $\{\sigma_u, \sigma_v\}$ span the tangent space of $S_{t_0}$, we conclude that $\sigma_t = V$ is tangent to $S_{t_0}$, implying that $f(u_0,v_0,t_0) = 0$.  Hence $p \in \mathcal{F}$.  So, $\mathcal{C} \subseteq \sigma(\mathcal{F})$. To prove that $\sigma(\mathcal{F}) \subseteq \mathcal{C}$, the above argument can be applied in the reverse direction.
}
\hfill $\square$ 

\begin{figure}
 \centering
 \includegraphics[scale=0.75]{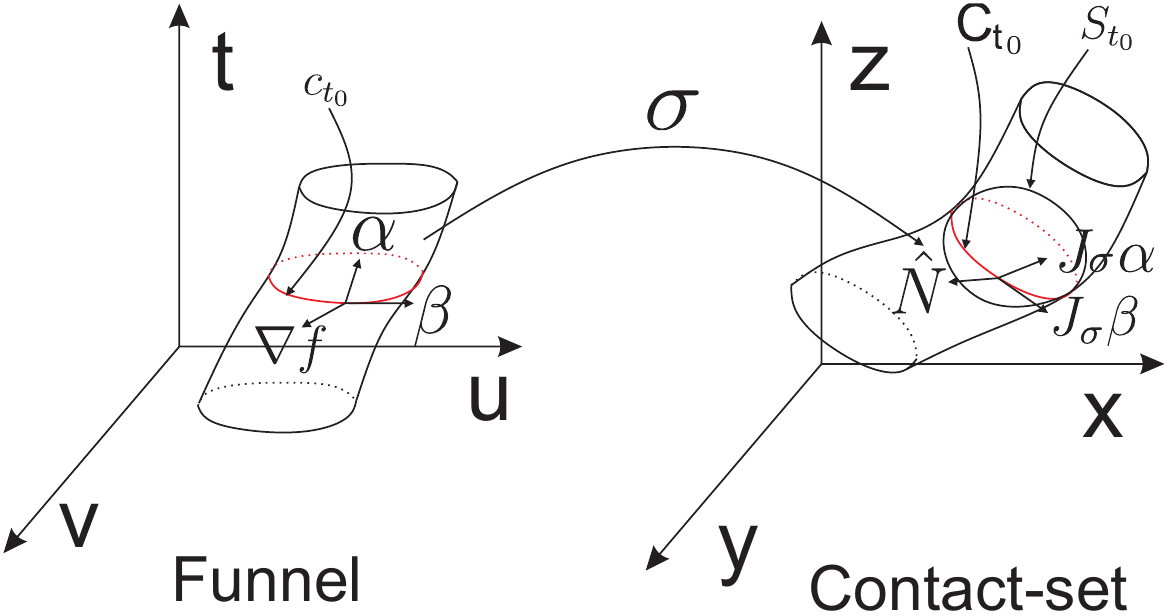}
 \caption{Funnel and contact-set}
 \label{funnelFig}
\end{figure}

Hence $S(u,v)$ `contributes' a point (namely, $A(t)S(u,v) + b(t)$) to the contact-set 
at time $t$ iff the triplet $(u,v,t)$ satisfies the following condition.
\begin{align} \label{envlCondEq}
f(u,v,t) = \left < V(u,v,t), \hat{N}(u,v,t) \right > = 0
\end{align}
\noindent  For a fixed $t$, Eq.~\ref{envlCondEq} is a system of one equation in two variables $u$ and $v$, hence, in a generic situation, the solution will be a curve.  
\eat{The union of such curves for all $t$ gives the final contact-set.}

Eq.~\ref{envlCondEq} can also be looked upon as the rank deficiency condition~\cite{jacobian} of the Jacobian $J_{\sigma}$ of the map $\sigma$ defined in~\ref{sweepDef}.
To make this precise, let
\begin{align} \label{jsigmaEq}
J_{\sigma} = 
\begin{bmatrix}
\sigma_u &  \sigma_v & \sigma_t
\end{bmatrix}_{3\times 3}
\end{align}
\noindent where $\sigma_u|_{(u,v,t)}= A(t)\frac{\partial S}{\partial u}(u,v)$ and $\sigma_v|_{(u,v,t)} = A(t)\frac{\partial S}{\partial v} (u,v)$ 
\eat{are the first order partial derivatives of $\sigma$ w.r.t. $u$ and $v$,} and $\sigma_t|_{(u,v,t)} = V(u,v,t)$.
Observe that regularity of $S$ ensures that $J_{\sigma}$ has rank at least 2. Further, it is easy to show that
$f(u,v,t)$ is a non-zero scalar multiple of the determinant of $J_{\sigma}$. 
Therefore, Eq.~\ref{envlCondEq} is precisely the rank deficiency condition of the Jacobian of $\sigma$.

\eat{The image of the map $\sigma$ restricted to the funnel $\mathcal{F}$ is the contact-set $\mathcal{C}$.
Note that $\mathcal{F}$ and $\mathcal{C}$ are both, in generic situation, two-dimensional manifolds and} 

Note that, for a point $(u_0,v_0,t_0) \in \mathbb{R}^3$, the Jacobian 
$J_{\sigma}|_{(u_0,v_0,t_0)}$ is a map from 
the tangent space to the ambient parameter-space at $(u_0,v_0,t_0)$ to the tangent space to the ambient object space at $\sigma(u_0,v_0,t_0)$.  
As already noted, if $(u_0,v_0,t_0) \in \mathcal{F}$, then $J_{\sigma}|_{(u_0,v_0,t_0)}$ is rank-deficient and maps 
the 3-dimensional ambient tangent space at $(u_0,v_0,t_0)$, (surjectively) onto, a 2-dimensional subspace of the ambient tangent space at $\sigma(u_0,v_0,t_0)$. \eat{by $J_{\sigma}|_{(u_0,v_0,t_0)}$.}  

The subset of $\mathcal{F}$ for a fixed value of $t$ will in general be a curve and will be referred to as the \emph{pcurve-of-contact} at time $t$ and its image through $\sigma$ will be a  subset of $\mathcal{C}$ which will be referred to as the \emph{curve-of-contact} since it is essentially the set of points on the surface $S$ where $S$ makes tangential contact with $\mathcal{C}$ at time $t$. The union of such curves-of-contact for all $t$ gives the contact-set $\mathcal{C}$.  The curve-of-contact at $t$ will be denoted by $C_t$ and the pcurve-of-contact will be denoted by $c_t$.  Fig.~\ref{funnelFig} schematically illustrates the funnel and the contact-set.  

\eat{Before we begin the discussion on smoothness of $\mathcal{F}$ and $\mathcal{C}$,}
Before proceeding further, we make the following the non-degeneracy assumption\footnote{Examples where this does not hold are (i) a cylinder being swept along its axis (ii) a planar face being swept in a direction orthogonal to its normal.  Such cases can be separately and easily handled.} that:
\begin{align}
\forall p \in \mathcal{F}, \nabla f|_{p}  \neq (0, 0, 0)	\label{nonDegAssEq}
\end{align}
Further, for ease of discussion, we assume that (i) $\mathcal{F}$ is connected, and (ii) $\forall (u,v,t) \in \mathcal{F}, (f_u,f_v) \neq (0,0)$
Our analysis can be easily extended to situations where these further assumptions do not hold.
\eat{See Section~\ref{conclusion} for situations where these assumptions do not hold.}

An important consequence of the assumption~\ref{nonDegAssEq} is that $\mathcal{F}$ is a 2-dimensional 
\eat{differentiable} manifold and, hence, $\mathcal{T}_{\mathcal{F}}(p)$ is 2-dimensional at all points $p \in \mathcal{F}$. Observe 
that $\mathcal{F}$ is also orientable as $\nabla f$ provides a continuous non-vanishing normal. 

Thus, $\mathcal{F}$ is topologically nice and regular, However, quite often, $\mathcal{C}$ has `anomalies' which arise due to self-intersections. 
One of the main contributions of this paper is a subtle, efficiently computable  mathematical function on $\mathcal{F}$ which allows to
identify points on $\mathcal{F}$ which give rise to anomalies in $\mathcal{C}$.

The key to our analysis ahead, is the restriction of the sweep map $\sigma$ to $\mathcal{F}$. We will abuse the notation, denote this
restriction, again by $\sigma$. So, $\sigma : \mathcal{F} \rightarrow \mathcal{C}$. 
Now, fix a point $p=(u,v,t) \in \mathcal{F}$ and let $q=\sigma(p) \in \mathcal{C}$. 
Since $det(J_{\sigma} (p))=0$, $\{\sigma_u(p), \sigma_v(p), \sigma_t(p)\}$ are linearly dependent.
As $S$ is regular, the set $\{\sigma_u(p), \sigma_v(p)\}$ forms a basis for the tangent space to $S_t$.
Therefore, we must have $\sigma_t = l\sigma_u +m \sigma_v $ where $l$ and $m$ are well-defined (unique) and are themselves
continuous functions of $u, v$ and $t$.

Clearly, $(\sigma_u(p), \sigma_v(p))$ is a natural (ordered) 2-frame in the object space at point $q$ (recall that, $q=\sigma(p)$).
Further, let ${\cal X}(p)$ be any ordered continuous 2-frame (basis) of the tangent space ${\mathcal{T}}_{\mathcal{F}}(p)$. Note
that, this 2-frame is in the parameter space and is associated to the point $p$. Now, through $\sigma$, more precisely, $J_{\sigma}(p)$,
the frame ${\cal X}(p)$ can be transported to another natural 2-frame $\sigma({\cal X}(p))$ in the object space at 
the point $q$. The determinant of  the linear transformation connecting these two natural frames at $q$, namely (i) $(\sigma_u(p) ,\sigma_v(p))$ 
and (ii) $\sigma({\cal X}(p))$ is the key to the subsequent analysis.
As we show later, this determinant is a positive scalar multiple of the continuous function $\theta: \mathcal{F} \rightarrow \mathbb{R}$ defined as
follows.
\eat{We define the continuous function $\theta :\mathcal{F} \rightarrow \mathbb{R}$ as follows:}
\begin{align}	\label{thetaEq}
 \theta (p)= l f_u +m f_v -f_t 
\end{align}
Here $p=(u,v,t)$ and $f_u, f_v$ and $f_t$ denote partial derivatives of the function $f$ w.r.t. $u,v$ and $t$ respectively at $p$, and $l$ and $m$
are as defined before.  Note that $\theta $ is easily and robustly computed.

We state an important result which we will prove in the coming sections:
\begin{thrm} 	\label{mainThm}
The function $\theta $ is such that (i) $\theta (p)<0$ indicates 
that $p$ is a point of local self intersection as defined by most authors, (see ~\cite{trimming,selfIntersections}) 
and (ii) $\theta (p)=0$ is where the rank of $J_{\sigma } (\mathcal{T}_\mathcal{F} (p))<2$, and finally (iii) 
excision of the region $\{ p | \theta (p)\leq 0 \}$ from the funnel $\mathcal{F}$ simplifies the construction of the envelope.
\end{thrm}

\eat{The function $\theta $ arises as a positive multiple of the determinant of 
the linear transformation connecting two natural $2$-frames on the 
tangent space ${\cal T}_\mathcal{C}$ of the contact set, 
namely (i) $(\sigma_u ,\sigma_v )$ and 
(ii) $\sigma ({\cal X})$, where ${\cal X}$ is any $2$-frame on ${\cal T}_{\cal F}$.
}

\subsection{A particular frame for $\mathcal{T}_{\mathcal{F}}$}  \label{funnelBasisSubSec}
Let $p = (u,v,t) \in \mathcal{F}$. In this section, we compute a natural 2-frame ${\cal X}(p)$ in $\mathcal{T}_{\mathcal{F}}(p)$. 
Note that, $\mathcal{F}$ being the zero level-set of the function $f$ defined in Eq.~\ref{envlEq}, $\nabla f|_p \bot \mathcal{T}_{\mathcal{F}}(p)$.  
We set $\beta = (-f_v, f_u, 0) \neq 0$ and note that $\beta \bot \nabla f$. It is easy to see that 
$\beta $ is tangent to the pcurve-of-contact $c_t$.  
Let $\alpha = \nabla f \times \beta  = (-f_uf_t, -f_vf_t, f_u^2+f_v^2)$.  
Here $\times$ is the cross-product in $\mathbb{R}^3$. Clearly, the set $\{ \alpha, \beta \}$ forms a basis of $\mathcal{T}_{\mathcal{F}}(p)$.

Figure~\ref{funnelFig} illustrates the basis $\{\alpha, \beta \}$ schematically. Observe that $\beta$ is tangent to the pcurve-of-contact at time $t$ 
and $\alpha$ points towards the `next' pcurve-of-contact.

\subsection{The determinant connecting the two frames}
We continue with the notation developed earlier. 
We have $\alpha = (-f_tf_u, -f_tf_v, f_u^2+f_v^2 )$ and $\beta = (-f_v, f_u, 0)$.  Hence,
\begin{align*}
J_{\sigma}\alpha &= -f_t f_u \sigma_u - f_t f_v \sigma_v + (f_u^2 + f_v^2) \sigma_t \\
			&=(-f_t f_u + l(f_u^2+f_v^2)) \sigma_u + (-f_t f_v + m(f_u^2+f_v^2))\sigma_v \\
J_{\sigma}\beta &= -f_v \sigma_u + f_u \sigma_v
\end{align*}
So, $\{ J_{\sigma}\alpha , J_{\sigma}\beta \}$ can be expressed in terms of $\{ \sigma_u, \sigma_v \}$ as follows
\begin{align*}
\begin{bmatrix}
J_{\sigma}\alpha & J_{\sigma}\beta
\end{bmatrix}
= 
\begin{bmatrix}
\sigma_u & \sigma_v
\end{bmatrix}
\underbrace{
\begin{bmatrix}
-f_t f_u + l(f_u^2+f_v^2) & -f_v \\
-f_t f_v + m(f_u^2+f_v^2) & f_u
\end{bmatrix} }_{\mathcal{D}(p)}
\end{align*}
Note that,
\begin{align}
det(\mathcal{D}(p)) &= (f_u^2 + f_v^2)(l f_u + m f_v - f_t) 	\label{detDcase1Eq} \\
&=(f_u^2 +f_v^2 ) \theta (p)  \label{thetaDet}
\end{align}

\subsection{Singularities of $\mathcal{C}$}
\begin{figure}
 \centering
 \includegraphics[scale=0.7]{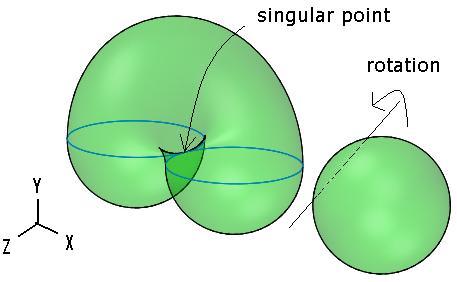}
 \caption{A sphere rotating about an axis which is tangent to itself.  There is type-1 L.S.I. but no type-2 L.S.I.}
 \label{type1Type2Fig}
\end{figure}

In this subsection, we propose an efficient test for detecting singularities on the contact-set.
See Fig.~\ref{type1Type2Fig} for an example.
Clearly, the detection of singularities is important in practice.

We start with the following definition.
\begin{defin} \label{lsiDef}
We say that the sweep causes a singularity if the composite map $\mathcal{F} \stackrel{\sigma}{\rightarrow} \mathcal{C} \hookrightarrow \mathbb{R}^3$ fails to be an immersion.  
\end{defin}
In other words, the sweep causes a singularity if $\exists p \in \mathcal{F}$ such that the rank of $J_{\sigma}(p)(\mathcal{T}_{\mathcal{F}}(p))$ is 
less than 2. Following the standard usage~\cite{diffTop}, in this case we say that the point $p$ is a critical point. 

\begin{lema} A point $r \in \mathcal{F}$ is a critical point iff $\theta(r)=0$ iff the rank of $J_{\sigma}(r)(\mathcal{T}_{\mathcal{F}}(r))$ is less than 2.
\end{lema}
\emph{Proof.}
By equation~\ref{thetaDet}, we have $\det(\mathcal{D}(r))=0$ iff $\theta(r)=0$. 
Recall that, as shown earlier, $\{\alpha(r), \beta(r)\}$ is a basis of $\mathcal{T}_{\mathcal{F}}(r)$,
and $\{\sigma_u(r), \sigma_v(r)\}$ is also a basis of the 2-frame associated at $\sigma(r)$. As $\mathcal{D}(r)$ is the matrix
expressing ${\cal{X}} = \{J_{\sigma}(r)(\alpha(r)), J_{\sigma}(r)(\alpha(r))$ in terms of $\{\sigma_u(r), \sigma_v(r)\}$, $\det(\mathcal{D}(r))=0$ iff
rank of $J_{\sigma}(r)(\mathcal{T}_{\mathcal{F}}(r))$ is less than 2. Thus, $r$ is a critical point iff $\theta(r)=0$ iff 
the rank of $J_{\sigma}(r)(\mathcal{T}_{\mathcal{F}}(r))$ is less than 2.
\hfill $\square$

Note that the above lemma proves part (ii) of theorem~\ref{mainThm}.

\begin{lema} A sweep causes a singularity if there exists points $p$ and $q$ on $\mathcal{F}$ such
that $\theta(p) \leq 0$ and $\theta(q) \geq 0$.
\end{lema}
\emph{Proof.} As $\theta$ is a continuous function on $\mathcal{F}$, the existence of $p$ and $q$ on $\mathcal{F}$
with the required properties implies existence of another point $r \in \mathcal{F}$ such that $\theta(r)=0$. 
By the previous lemma, this implies that the sweep causes a singularity.
\hfill $\square$

The above lemma leads to a computationally efficient test for detecting singularities: namely, evaluating $\theta$ at sampled points
on $\mathcal{F}$ and checking if it changes sign on $\mathcal{F}$.

\begin{figure*}
 \centering
 \includegraphics[scale=0.35]{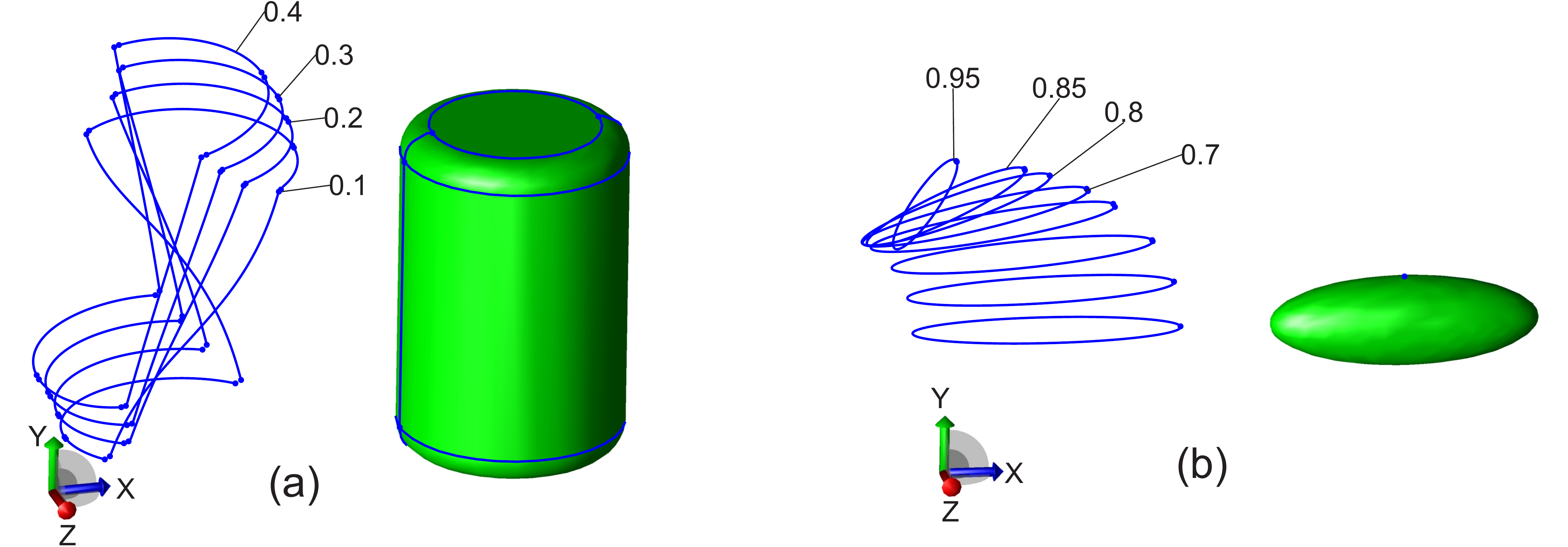}
 \caption{Examples of local self-intersection showing curves-of-contact at few time instances: (a) A cylinder with blended edges undergoing translation and rotation about $x$-axis (b) An ellipsoid undergoing translation along a curvilinear path 
 \eat{(c)An ellipsoid undergoing translation and rotation about $y$-axis (d)A sphere undergoing pure rotation about $z$-axis}}
 \label{lsiExamFig}
\end{figure*}

The analysis done so far helps us detect singularities on the contact-set.  In the next section we will 
perform a detailed analysis of local self-intersections which is topological in nature.
Towards this, note that all points in the non-critical set may not lead to points on the envelope of the swept volume.
For some $p=(u_0,v_0,t_0) \in \mathcal{F}$, $\sigma(p)$ may lie in the {\em interior} of the solid $M_t$ of which the surface patch $S_t$ is a part of,
for some $t$ in neighbourhood of $t_0$.  
In that case $\sigma(p)$ will not be on the envelope. Fig.~\ref{lsiExamFig} shows two sweeping examples with self-intersections. Curves-of-contact at a few time instances are shown. 
In the next section we focus on identifying such points.

\section{Topological and regularity analysis of ${\mathcal{C}}$} \label{lsiSec2}
\begin{figure}
 \centering
 \includegraphics[scale=0.5]{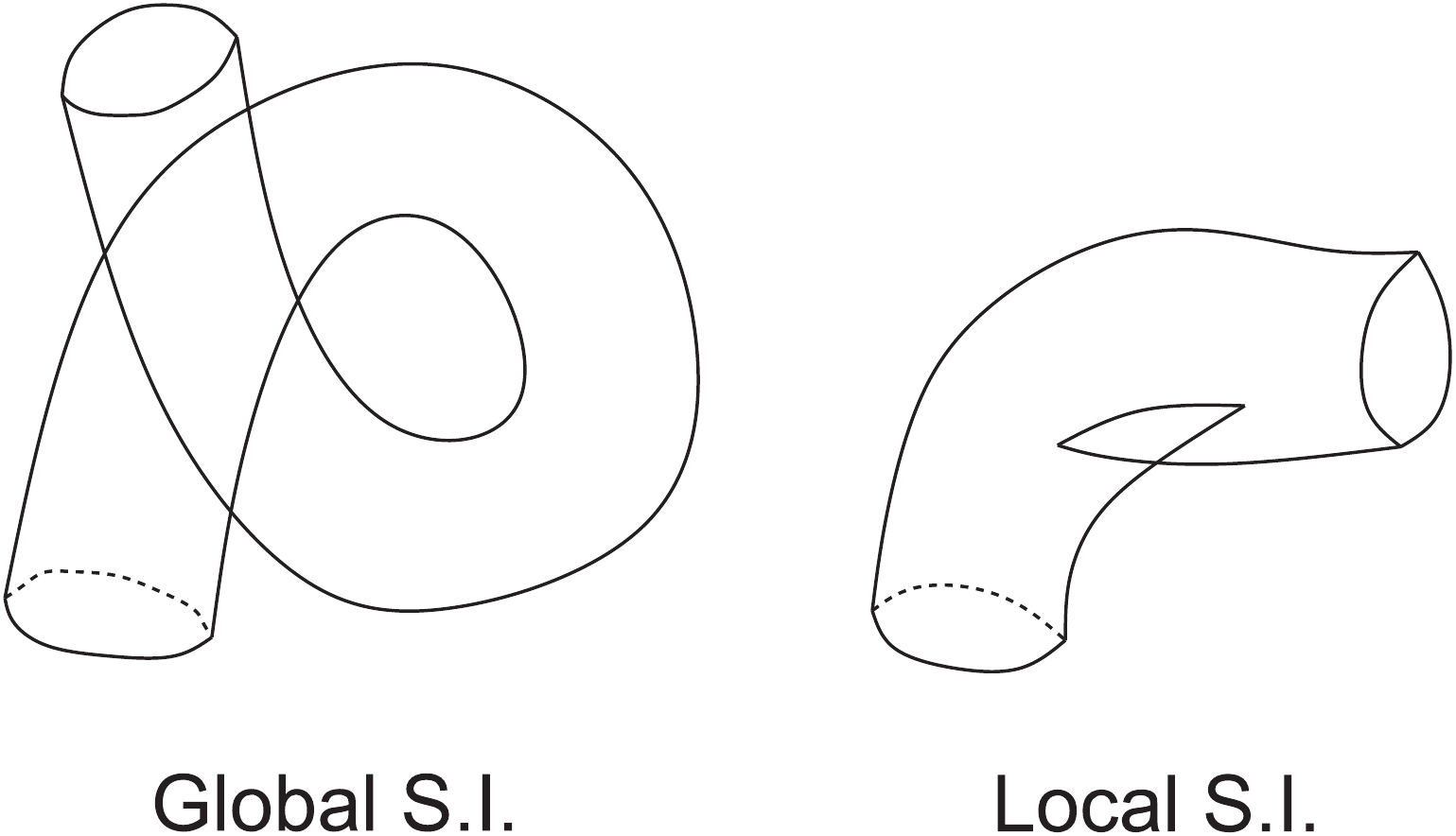}
 \caption{Global and local self-intersection}
 \label{globalLocalFig}
\end{figure}

Quite often, the anomalies on $\mathcal{C}$ arise due to self-intersections. 
If $\mathcal{C}$ has self-intersections, it needs to be trimmed to obtain the envelope of the swept volume~\cite{trimming}.  
Self-intersections can be broadly be classified into global and local. Fig.~\ref{globalLocalFig} illustrates the difference 
between global and local self-intersections schematically. 

If there are only global self-intersections occurring on $\mathcal{C}$, the composite map 
$\mathcal{F} \stackrel{\sigma}{\rightarrow} \mathcal{C} \hookrightarrow \mathbb{R}^3$ fails to be an injection.
However, it is an immersion (see~\cite{diffTop}), i.e. 
$\forall p \in \mathcal{F}$, the rank of $J_{\sigma}(p)({\mathcal{T}_{\mathcal{F}}(p)})$ is 2.
In principle, global self-intersections can be detected by surface-surface intersection. (see~\cite{procedural})

The case of local self-intersection is more subtle as it leads to singularities on $\mathcal{C}$. 
Clearly, the detection of local self-intersections is also central to a robust implementation of the solid sweep in CAD systems.

In literature~\cite{trimming,selfIntersections}, local self-intersections have been quantified by looking at points in the contact-set which lie in the interior of the solid $M_t$ for 
some time instant $t$.  Clearly, such a point cannot be on the envelope of the swept volume.  This approach was used in~\cite{trimming} for detecting local 
and global self-intersections, where the authors used implicit representation of the surface bounding the solid which is being swept. 
We adapt this concept to when the surface of the solid is represented parametrically.  
We will refer to this type of local self-intersection as \emph{type-2 L.S.I.}. It turns out that type-2 L.S.I. is intimately related to
the analysis carried out in the previous section. To make this connection precise, we first introduce another type of local self-intersection.
For lack of a better name, this is called as \emph{type-1 L.S.I.}

\subsection{Type-1 local self-intersections}  \label{type1SubSec}

\begin{defin} \label{type1Def}
A type-1 L.S.I. is said to occur at a point $p \in \mathcal{F}$ if  $\theta(p) \leq 0$.
\end{defin}

Thus, Type-1 L.S.I is {\em our} classification of a local self intersection. 
We will see in subsection~\ref{lsiRelationSubSec} that a for a Type-1 L.S.I 
point $p$, the image  
$\sigma(p)$ does not lie on the envelope of the swept volume.

\subsection{Type-2 local self-intersection}  \label{type2SubSec}
In order to define type-2 L.S.I. we first describe the inverse trajectory corresponding to a given trajectory~\cite{trimming, classifyPoints}.

Given a trajectory as in definition~\ref{trajectoryDef} and a fixed point $x$ in object-space, we would like to compute the set of points in the object-space which get mapped to $x$ at some time instant.  This set can be computed through the inverse trajectory defined as follows.

\begin{defin} \label{invTrajDef}
Given a trajectory $h$, the \emph{inverse trajectory} $\bar{h}$ is defined as the map $\bar{h}:[0,1]  \to (SO(3), \mathbb{R}^3)$ given by $\bar{h}(t) = (A^t(t), -A^t(t)b(t))$.
\end{defin}

Thus, for a fixed point $x \in \mathbb{R}^3$, the inverse trajectory of $x$ is the map $\bar{y}:[0,1] \to \mathbb{R}^3$ given by $\bar{y}(t) = A^t(t)(x - b(t))$. The range of $\bar{y}$ is $\{ A^t(t)x - A^t(t)b(t) | t \in [0,1] \} = \{ z \in \mathbb{R}^3 | \exists t \in [0,1], A(t)z + b(t) = x\}$.  We will denote the trajectory of $x$ by $y:[0,1] \to \mathbb{R}^3$, $y(t) = A(t)x + b(t)$.  We now note a few useful facts about inverse trajectory of $x$.  We assume without loss of generality that $A(t_0) = I$ and $b(t_0)=0$.  Denoting the derivative with respect to $t$ by $\dot{}$, we have
\begin{align}
\dot{\bar{y}}(t)=\dot{A}^t(t)(x-b(t)) - A^t(t)\dot{b}(t)	\label{yBarDotEq}
\end{align}
Since $A \in SO(3)$ we have,
\begin{align}
 A^t(t) A(t) &= I, \forall t 	\label{aSO3Eq}
\end{align}
Differentiating Eq.~\ref{aSO3Eq} w.r.t. $t$ we get
\begin{align}	
\dot{A}^t(t) A(t) + A^t(t)\dot{A}(t) &= 0, \forall t 	\label{aDotEq}  \\
\dot{A}^t(t_0) + \dot{A}(t_0) &= 0			\label{aDotTNotEq}
\end{align}
Differentiating Eq.~\ref{aDotEq} w.r.t. $t$ we get
\begin{align}
\nonumber \ddot{A}^t(t)A(t) + 2\dot{A}^t(t)\dot{A}(t) + A^t(t)\ddot{A}(t) &=0, \forall t  \\
\ddot{A}^t(t_0) + 2\dot{A}^t(t_0)\dot{A}(t_0) + \ddot{A}(t_0) &=0		\label{aDotDotTNotEq}
\end{align}
Using Eq.~\ref{yBarDotEq} and Eq.~\ref{aDotTNotEq} we get
\begin{align}
\dot{\bar{y}}(t_0) &= -\dot{A}(t_0)x - \dot{b}(t_0) = -\dot{y}(t_0)		\label{yBarDotTNotEq} 			
\end{align}
Differentiating Eq.~\ref{yBarDotEq} w.r.t. time we get
\begin{align}
\ddot{\bar{y}}(t) = \ddot{A}^t(t)(x-b(t)) - 2\dot{A}^t(t)\dot{b}(t) - A^t(t)\ddot{b}(t) 	\label{yBarDotDotEq}
\end{align}
Using Equations~\ref{yBarDotDotEq}, \ref{aDotTNotEq} and \ref{aDotDotTNotEq} we get
\begin{align}
 \ddot{\bar{y}}(t_0) = -\ddot{y}(t_0) + 2\dot{A}(t_0)\dot{y}(t_0)		\label{yBarDotDotTNotEq}
\end{align}
We now define type-2 L.S.I.  The surface $S$ is the boundary of the solid $M$ being swept.  We will refer to the interior of $M$ by $Int(M)$ and exterior of $M$ by $Ext(M)$.


\begin{defin} \label{type2Def}
A \emph{type-2 L.S.I} is said to occur at a point $(u_0, v_0, t_0)$ if the inverse trajectory of the point $\sigma(u_0, v_0, t_0)$ intersects $Int(M_{t_0})$.(see~\cite{trimming})
\end{defin}

\begin{figure}
 \centering
 \includegraphics[scale=0.8]{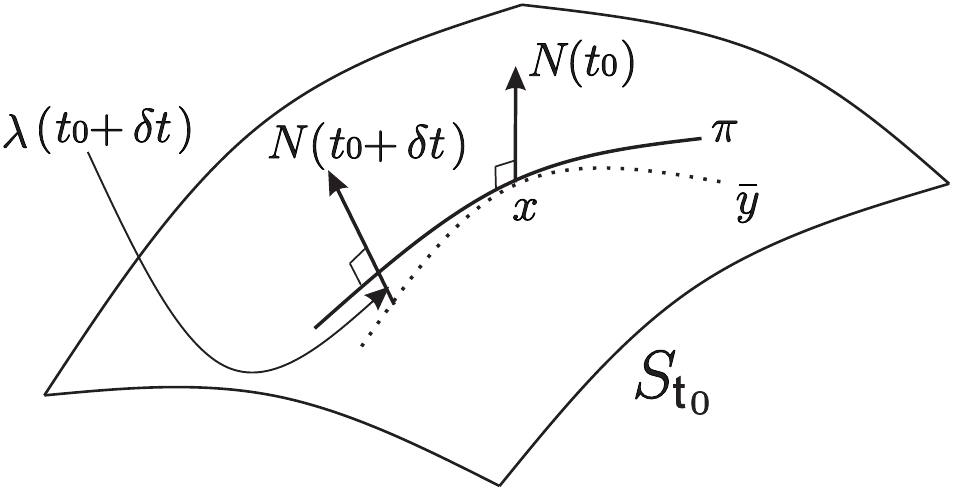}
 \caption{Type-2 local self-intersection}
 \label{type2LSIFig}
\end{figure}

Fig.~\ref{type2LSIFig} illustrates type-2 L.S.I. schematically where $\bar{y}$ is the inverse trajectory of the point $x \in S_{t_0}$ and $\pi$ is the projection of $\bar{y}$ on $S_{t_0}$. Suppose the point $x = \sigma(u_0,v_0,t_0) \in \mathcal{C}$.  Let $\lambda(t)$ be the signed distance of $\bar{y}(t)$ from surface $S_{t_0}$.  If the point $\bar{y}(t)$ is in $Int(M_{t_0})$, $Ext(M_{t_0})$ or on the surface $S_{t_0}$, then $\lambda(t)$ is negative, positive or zero respectively.  Then we have $\bar{y}(t) - \pi(t) = \lambda(t)N(t)$, where $\pi(t)$ is the projection of $\bar{y}(t)$ on $S_{t_0}$ along the unit outward pointing normal $N(t)$ to $S_{t_0}$ at $\pi(t)$.  Then, the following relation holds for $\lambda$.
\begin{align} \label{lambdaEq}
\lambda(t) = \left < \bar{y}(t) - \pi(t) , N(t) \right >
\end{align}

We now give a necessary and sufficient condition for type-2 L.S.I. to occur.

\begin{lema} \label{type2Lem}
Type-2 L.S.I. occurs at a point $p=(u_0,v_0,t_0) \in \mathcal{F}$ if and only if either of the following conditions hold
\begin{enumerate}
\item $\ddot{\lambda}(t_0)= \left < -\ddot{\sigma}+2\dot{A}V , N \right > + \kappa v^2 < 0$ where $\kappa$ is the normal curvature of $S_{t_0}$ at $(u_0,v_0)$ along velocity $V(p)$, $N$ is the unit length outward pointing normal to $S_{t_0}$ at $(u_0,v_0)$ and $v^2 = \left< V(p), V(p) \right>$.
\item $\lambda(t)$ is negative for some $t$ in some nbd of $t_0$.
\end{enumerate}
\end{lema}
{\bf Remark}: The statement of the above lemma (except for the insightful expression of $\ddot{\lambda}(t_0)$) is similar in spirit to the key Theorem-2 in~\cite{trimming} in the context of implicitly defined solids. 

\noindent \emph{Proof.} Differentiating Eq.~\ref{lambdaEq} with respect to time and denoting derivative w.r.t. $t$ by $\dot{}$, we get
\begin{align}
  \dot{ \lambda}(t) &= \left < \dot{\bar y}(t) - \dot{\pi}(t), N(t) \right > + \left < \bar{y}(t) - \pi(t) , \dot{N}(t) \right > \\
\nonumber  \ddot{\lambda}(t) &= \left < \ddot{\bar{y}}(t) - \ddot{\pi}(t), N(t) \right > + 2\left < \dot{\bar y}(t) - \dot{\pi}(t), \dot{N}(t) \right > \\
		&+ \left < \bar{y}(t) - \pi(t), \ddot{N}(t) \right >	\label{ddotLambdaEq}
\end{align}
At $t=t_0$, $\bar{y}(t_0) = \pi(t_0)$.  Since $ \dot{y}(t_0)=V(p) \bot N(p)$, it follows from Eq.~\ref{yBarDotTNotEq} that $\dot{\bar{y}}(t_0) \bot N(p)$.  It is easy to verify that $\dot{\pi}(t_0) = \dot{\bar{y}}(t_0)$.  Hence, 
\begin{align}
\lambda(t_0) = \dot{\lambda}(t_0) = 0 \label{lambdaTNotEq}
\end{align}
From Eq.~\ref{ddotLambdaEq} and Eq.~\ref{yBarDotDotTNotEq} it follows that
\begin{align}
\nonumber \ddot{\lambda}(t_0) &= \left < \ddot{\bar{y}}(t_0) - \ddot{\pi}(t_0), N(t_0) \right >\\
					&= \left < -\ddot{y}(t_0) + 2\dot{A}(t_0)\dot{y}(t_0) - \ddot{\pi}(t_0), N(t_0) \right >  \label{ddotLambdaTNotEq}
\end{align}
Since $\pi(t) \in S_{t_0} \forall t$ in some neighbourhood $U$ of $t_0$, we have that $\left < \dot{\pi}(t), N(t) \right > = 0, \forall t \in U$.  Hence $\left < \ddot{\pi}(t), N(t) \right> + \left < \dot{\pi}(t), \dot{N}(t) \right > = 0, \forall t \in U$.  Hence $-\left < \ddot{\pi}(t_0), N(t_0) \right > =  \left< \dot{\pi}(t_0), \dot{N}(t_0) \right > =  \left< \dot{\pi}(t_0), \mathcal{G}^*(\dot{\pi}(t_0)) \right > = \left < \dot{y}(t_0), \mathcal{G}^*(\dot{y}(t_0)) \right>$ = $\left < V(p) , \mathcal{G}^*(V(p)) \right > =\kappa v^2$.  Here $\mathcal{G}^*(\dot{y})$ is the differential of the Gauss map, i.e. the curvature tensor of $S_{t_0}$ at point $x$.  Using this in Eq.~\ref{ddotLambdaTNotEq} and the fact that  $\dot{y}(t_0) = \dot{\sigma}(p)$, $\ddot{y}(t_0) = \ddot{\sigma}(p)$ we get
\begin{align}
\ddot{\lambda}(t_0)  &= \left < -\ddot{\sigma}(p) + 2\dot{A}(t_0)V(p) , N(t_0) \right > + \kappa v^2  \label{lsi2Eq}
\end{align}
From Eq.~\ref{lambdaTNotEq} and Eq.~\ref{lsi2Eq} we conclude that if $\ddot{\lambda}(t_0) < 0$ the point $x=\bar{y}(t_0) = \sigma(u_0,v_0,t_0)$ is a local maxima of the function $\lambda$ and the inverse trajectory of $x$ intersects with interior of the solid $M_{t_0}$ causing type-2 L.S.I.  Similarly, if $\ddot{\lambda}(t_0) > 0$ we conclude that $x$ is a local minima of $\lambda$ and the inverse trajectory of $x$ does not intersect with the interior of $M_{t_0}$ and there is no L.S.I. occurring at $x$.  

However, if $\ddot{\lambda}(t_0)$ is zero, one needs to inspect a small neighbourhood of $t_0$ to see if $\exists t$ such that $\lambda(t)<0$ in order to check for type-2 L.S.I.
\hfill $\square$

If $\ddot{\lambda} = 0$ at a point, the structure of the contact-set $\mathcal{C}$ is unknown at that point.  We will see in the next subsection that at such a point, $\mathcal{C}$ has singularity.

\subsection{Relation between type-1 L.S.I. and type-2 L.S.I.} \label{lsiRelationSubSec}

In this subsection we will see that type-2 L.S.I. implies type-1 L.S.I. at any point $p = (u_0,v_0,t_0) \in \mathcal{F}$ 

\begin{lema} \label{type1Type2Lem}
$\theta(p) = \ddot{\lambda}(t_0)$. \eat{In particular, $sign(det(\mathcal{D}(p))) = sign(\ddot{\lambda})(t_0)$.}
\end{lema}
\emph{Proof.}
\eat{
Using Eq.~\ref{detDcase1Eq} and the fact that $(f_u^2+f_v^2) > 0$ we see that $sign(det(\mathcal{D})) = sign(l f_u + m f_v - f_t)$.  Further,
}
Recalling definition of $\theta(p)$ from Eq.~\ref{thetaEq}
\begin{align*}
l f_u + m f_v - f_t &= \left< l\hat{N}_u + m\hat{N}_v, V\right> + \left<\hat{N}, l V_u + m V_v \right >\\  &-  \left< \hat{N}_t, V\right > - \left<\hat{N} ,V_t \right >
\end{align*}
Here $\hat{N}_u = \mathcal{G}^*(\sigma_u)$ and $\hat{N}_v = \mathcal{G}^*(\sigma_v)$ where $\mathcal{G}^*$ is the shape operator (differential of the Gauss map) of $S_{t_0}$ at $(u_0,v_0)$.  Also, $V_u = A_t S_u$ and $V_v = A_t S_v$.  Assume without loss of generality that $A(t_0) = I$ and $b(t_0) = 0$. Using Eq.\ref{aDotTNotEq} and the fact that $\sigma_t = l\sigma_u+m\sigma_v$  we get
\begin{align}
\nonumber l f_u + m f_v - f_t &= \left< \mathcal{G}^*V, V \right > + 2\left<A_t V ,N \right> - \left <V_t, N\right >  \\
			& = \kappa v^2  + \left < 2A_t V - V_t  , N \right > \label{lsiRelationEq}
\end{align}
From Eqs.~\ref{lsi2Eq} and~\ref{lsiRelationEq} and the fact that $\frac{\partial \sigma}{\partial t^2}=V_t$ we get 
$\theta(p) = l f_u + m f_v - f_t = \ddot{\lambda}$
\hfill $\square$

\begin{figure}
 \centering
 \includegraphics[scale=0.7]{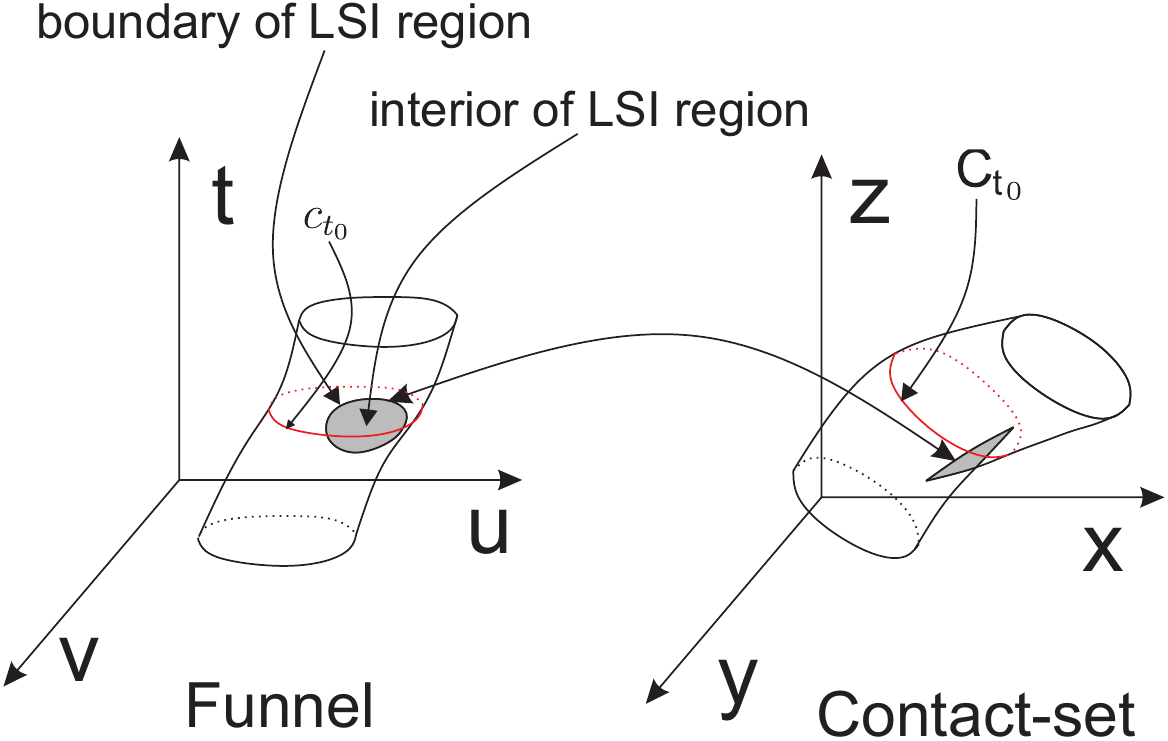}
 \caption{Region of local self-intersection}
 \label{lsiRegionFig}
\end{figure}
Let $p$ be such that $\theta(p) < 0$.  By lemma~\ref{type1Type2Lem}, $\ddot{\lambda}(t_0)<0$. Further, by lemma~\ref{type2Lem} a type-2 L.S.I. occurs at $p$. This proves parts (i) and (iii), and hence completes the proof of theorem~\ref{mainThm}.

Fig.~\ref{lsiRegionFig} schematically illustrates the region on the funnel where local self-intersection occurs and the corresponding region on the contact-set.  
A curve-of-contact and the corresponding pcurve-of-contact is shown in red colour at a time instant $t_0$ when a local self-intersection occurs. The shaded region there
corresponds to $\theta < 0$. Thus $sign(\theta)$ 
changes(from $-$ve to $+$ve) as one moves from the interior to the exterior of the shaded region. 
Of course $\theta=\ddot{\lambda} = 0$ on the boundary of the shaded region where $\mathcal{C}$ has a singularity.

Fig.~\ref{type1Type2Fig}  shows an example of sweeping which clearly demonstrates the subtle difference between type-1 L.S.I. and type-2 L.S.I. 
\eat{Type-1 L.S.I. occurs at time $t_0$ if the neighbourhood of $t_0$, $\exists t$ such that either $C_{t_0} \cap C_{t} \neq \emptyset$ or $C_{t_0} \cap Int(M_t) \neq \emptyset$.  On the other hand, type-2 L.S.I. occurs at time instant $t_0$ if in the neighbourhood of $t_0$, $\exists t$ such that $C_{t_0} \cap Int(M_t) \neq \emptyset$, where $Int(M_t)$ denotes the interior of the solid $M_t$.}
In Fig.~\ref{type1Type2Fig} a sphere is rotated about an axis which is tangent to itself.  The contact-set is shown in green and the curve-of-contacts on the contact-set at initial position $(t = 0)$ and final position $(t = 1)$ are also shown in blue. In Fig.~\ref{type1Type2Fig} there is type-1 L.S.I. but no type-2 L.S.I. since no 
point of the contact-set intersects with the interior of solid at any time.

We now present two examples of sweeping which result in local self-intersections.  We apply the test for type-1 L.S.I. on these examples to demonstrate its effectiveness.  \eat{We show the computation of $det(\mathcal{D})$ for the first example.}

\begin{exmpl} \label{cylinderExam}
Consider a solid cylinder along the $y$-axis with height $2.5$ and radius $2$, parameterized as $S(u, v) = (2\cos v, u, -2\sin v)$, $u \in [-1.25, 1.25]$, $v \in [-\pi, \pi]$ being swept along the trajectory given by $h(t) = (A(t), b(t))$ where $A(t) = \begin{bmatrix}1 & 0 & 0 \\0 & \cos(0.1 \pi t) & -\sin(0.1 \pi t) \\ 0 & \sin(0.1 \pi t) & \cos(0.1\pi t)  \end{bmatrix}$ and $b(t) = \begin{bmatrix} 3 \cos\left (\frac{\pi}{2}t\right ) -3\\ 3 \sin \left (\frac{\pi}{2}t \right ) \\ 0  \end{bmatrix}$ $t \in [0,1]$.  The resulting envelope has local self-intersections as illustrated in Fig.~\ref{lsiExamFig}(a) which shows the cylinder with blended edges and curves-of-contact at few time instances.  Type-1 L.S.I is detected at time $t = 0.1$  by the test given in lemma~\ref{type2Lem} at point $p = (u=0.18, v=1.53, t=0.1)$. 
\eat{ At $p$, $\hat{N} = \begin{bmatrix}-0.037\\ -0.037\\0.998\end{bmatrix}, V =\begin{bmatrix} -0.737\\ 5.278\\0.138 \end{bmatrix}, J_{\sigma}= \begin{bmatrix} 0.0 & -1.998 & -0.737 \\1.999 & 0.002 & 5.278 \\0.062 &-0.075 & 0.138 \end{bmatrix}, \nabla f = \begin{bmatrix}0.482\\-0.883\\-1.414  \end{bmatrix}, \alpha = \begin{bmatrix}0.686\\-1.258\\1.012 \end{bmatrix}, \beta=\begin{bmatrix} 0.883\\0.482\\0.0 \end{bmatrix} $ and}
 $\theta(p) =  -2.378$.
\end{exmpl}

\begin{exmpl} \label{ellipTranExam}
Consider an ellipsoid with axes lengths $3$, $1$, and $1$ parameterized as $S(u,v) = (-3\cos(u) \cos(v), \sin(u),\\ \cos(u) \sin(v))$, $u \in [-\frac{\pi}{2}, \frac{\pi}{2}], v \in [-\pi, \pi]$ being swept along the curvilinear trajectory given by $h(t) = (A(t), b(t))$ where $A(t) = I \forall t$ and $b(t) = \begin{bmatrix} 3 \cos\left (\frac{\pi}{2}t\right ) -3\\ 3 \sin \left (\frac{\pi}{2}t \right ) \\ 0  \end{bmatrix}$ $t \in [0,1]$. The curves-of-contact are shown in Fig.~\ref{lsiExamFig}(b).  Type-1 L.S.I. is detected at time $t=0.8$.  $\theta(p) =  -11.864$ at $(-0.791, -0.157, 0.8)$. 
\end{exmpl}

\eat{
\begin{exmpl} \label{ellipTranRotExam}
Consider the ellipsoid described in example~\ref{ellipTranExam} being swept along the trajectory given by $h(t) = (A(t), b(t))$ where $A(t) = \begin{bmatrix}\sin(0.1 \pi t) & 0 & \cos(0.1\pi t) \\0 & 1 & 0 \\ \cos(0.1\pi t) & 0 & -\sin(0.1\pi t)  \end{bmatrix}$ and $b(t) = \begin{bmatrix} 3 \cos\left (\frac{\pi}{2}t\right ) -3\\ 3 \sin \left (\frac{\pi}{2}t \right ) \\ 0  \end{bmatrix}$ $t \in [0,1]$.  This is illustrated in Fig.~\ref{lsiExamFig}(c).  Type-1 and type-2 L.S.I. are detected at time $t=0.85$.  $det(\mathcal{D}) =  -14.74$ at $(-0.958,-0.153, 0.85)$. 
\end{exmpl}

\begin{exmpl} \label{sphRotExam}
Consider a sphere with radius $3$ and center $(2, 0, 0)$ parameterized as $S(u,v) = (2-3\cos(u) \cos(v)\\, 3\sin(u), 3\cos(u) \sin(v))$, $u \in [-\frac{\pi}{2}, \frac{\pi}{2}], v \in [-\pi, \pi]$ undergoing pure rotation about $z$-axis.  The trajectory is given by $h(t) = (A(t), b(t))$ where $A(t) = \begin{bmatrix} \cos(\pi t) & -\sin(\pi t) & 0 \\ \sin(\pi t) & \cos(\pi t) & 0 \\ 0 & 0 & 1 \end{bmatrix}$ and $b(t) = 0$, $t \in [0,1]$.  The curves-of-contact are shown in Fig.~\ref{lsiExamFig}(d). Type-1 and type-2 L.S.I. are detected at time $t=0$.  $det(\mathcal{D}) =  -250.16$ at $(0.0,-0.157, 0.0)$.
\end{exmpl}
}


\section{Mathematical structure of the smooth case} \label{smoothSec}
In this section, we consider the smooth case where the solid $M$ is composed of faces meeting smoothly. 
As usual, each face (or the associated surface patch) is smooth (of class $C^k$ for $k \geq 2$). Further,
adjacent faces meet smoothly at the common edge. This is referred to as $G^1$ continuity~\cite{farin} which formally
means that the unit outward normals to the adjacent faces match on the common edge. Similarly, 
at a vertex, all the unit outward normals to faces incident on this vertex are identical. 
The solid shown in figure~\ref{brepFig} is such a solid.

Consider a sweep of $M$ along a trajectory $h$ which causes no self-intersections/anomalies on the contact-set $\mathcal{C}$.
As described in Section~\ref{architecture}, every point $p$ on $\mathcal{C}$ comes from a curve of contact on $M$ and therefore
is associated to a point $q$ of $M$. 
Let $\pi: \mathcal{C} \rightarrow M$ ($p \mapsto \pi(p)=q$) denote this natural map.
For every $p \in \mathcal{C}$, $\pi(p)$ belongs to some geometric entity of $M$, i.e., a vertex, edge or face.  
This sets up the natural correspondence between
geometric entities of $\mathcal{C}$ and that of $M$. For a face $F$ of $M$, let $\mathcal{C}_F$ denote the part of $\mathcal{C}$ which corresponds
to the face $F$ under this correspondence. For example, in figure~\ref{sweepExample}, the green face on the solid $M$ corresponds
to multiple green faces of $\mathcal{C}$. Clearly, the map $\pi$ restricts naturally from $\mathcal{C}_F \rightarrow F$.

There are situations in which, for example, an edge (or a part of it) on $M$ remains
on the boundary for a while and thus, `sweeps' a face on $\mathcal{C}$. For simplicity, we assume that such situations are ruled out.
In other words, no lower dimensional geometric entity of $M$ gives rise to a higher dimensional geometric entity on $\mathcal{C}$. 
This is the case for the sweep operation illustrated in figure~\ref{sweepExample}.

\subsection{Local similarity within a face}
\eat{In this subsection, we show that the topology of a face of $\mathcal{C}$ mimics locally that of the corresponding face of the solid $M$.}
Firstly, recall that each face $F$ of $M$ is derived from  an underlying surface $S_F$ by restricting the 
parameters of $S_F$ to a suitable domain $D_F$. 
Now, by applying the `local' analysis of Section~\ref{lsiSec} to the surface $S_F$, we have

\begin{lema}\label{localnormallemma}
The set $\mathcal{C}_F$ has no self-intersections and is a smooth manifold. Further, let $p \in \mathcal{C}_F$ correspond to $q \in F$ at time $t$.
Then, the unit normal $N(p)$ to $\mathcal{C}_F$ at $p$ is simply $A(t)N(q)$ where $N(q)$ is the unit normal to $F$ at $q$.
\end{lema}

Further, we would like to show that `locally', $\mathcal{C}_F$ has the same topology as that of $F$.
More precisely, the natural map $\pi: \mathcal{C}_F \rightarrow F$ is a local homeomorphism onto its image.
Thanks to the `local' nature, we may analyse this via the underlying surface $S_F$. For ease of notation, we sometimes omit the reference to $F$
and freely use notations from Section~\ref{lsiSec}. As the sweep $\sigma$ is free of self-intersections, the key map 
$\sigma: \mathcal{F} \rightarrow \mathcal{C}_F$ is a bijective immersion (recall that, in Section~\ref{lsiSec}, $\mathcal{F}$ is the funnel defined
via the function $f=0$). Thus, by the inverse function theorem, it is invertible via a continuous inverse. Therefore, in order to show that 
$\pi: \mathcal{C}_F \rightarrow F$ is a local homeomorphism onto its image it suffices to prove the following lemma.

\eat{
\begin{lema} \label{homeoLem}
Let $p=(u_0,v_0,t_0) \in \mathcal{F}$. Then there exists a neighbourhood $\mathcal{M}$ of $S(u_0,v_0)$ in $S$ and a neighbourhood 
$\mathcal{P}$ of $\sigma(p)$ in $\mathcal{C}_F$ such that there exists a continuous bijective map from $\mathcal{M}$ to $\mathcal{P}$.
\end{lema}
\emph{Proof.}  
As in Section~\ref{lsiSec}, we make the simplifying assumption that $f_t \neq 0$ at $p$.
By the implicit function theorem, there exists a neighbourhood $\mathcal{N}$ of $(u_0, v_0)$ and a 
function $g(u,v)=t$ such that  $\forall (u,v) \in \mathcal{N}$, $f(u,v,g(u,v))=0$.  It follows that 
there exists a neighbourhood $\mathcal{M} = S(\mathcal{N})$ of point $S(u_0,v_0)$  in which $S(u,v) \mapsto A(t)S(u,v)+b(t)$ is a well defined map, 
where $t = g(u,v)$.  We thus define a map from the subset $\mathcal{M}$ of $S$ to its corresponding 
subset of $\mathcal{C}_F$ given by $\Gamma: \mathcal{M} \to \Gamma(\mathcal{M}) \subset \mathcal{C}_F$, 
$\Gamma(S(u,v)) = A(t)S(u,v)+b(t)$, where $t=g(u,v)$.  Note that if there are no self-intersections on 
$\mathcal{C}_F$ then the map $\Gamma$ is injective and $\Gamma^{-1}: \Gamma(\mathcal{M}) \to \mathcal{M}$ is well-defined.  
$S$ and $S^{-1}$ are both continuous maps.  Hence $\forall q \in \mathcal{M}$, 
$\Gamma(q) = A(g(S^{-1}(q)))q + b(g(S^{-1}(q)))$ is a composition of continuous maps and is continuous.  
\hfill $\square$
}

\begin{lema} \label{homeoLem}
The natural map $\pi': \mathcal{F} \rightarrow F$ defined as: for $p=(u,v,t) \in
\mathcal{F}$, $\pi'(p) = S(u,v)$, 
is a local homeomorphism. 
\end{lema}
{\bf Remark} The map $\pi'$ is simply the composition of $\sigma$ and $\pi$.
\emph{Proof.} Let $p=(u_0,v_0,t_0) \in \mathcal{F}$. Here, the assumption that $f_t
\neq 0$ at $p$ is very crucial.
Firstly, by the implicit function theorem, there exists a neighbourhood $O$ of
$(u_0, v_0)$ and a 
continuous function $t=g(u,v)$ defined on $O$  such that  $\forall (u,v) \in O$,
$f(u,v,g(u,v))=0$. Further, 
the set $O(p) = \{(u,v,g(u,v)) \mid (u,v) \in O\}$ is a neighbourhood of $p$ in
$\mathcal{F}$. 
On this neighbourhood of $p$, the function $\pi'$ is a bijection and invertible.
This easily follows from
the fact that, for every $(u,v) \in O$, there is a unique time $t$, namely,
$t=g(u,v)$, such that $(u,v,g(u,v)) \in O(p)$.
\hfill $\square$

\eat{\begin{lema} More precisely, for a point $p \in \mathcal{C}_F$, there exists a nbd $D(p)$ of $p$ in $\mathcal{C}_F$ and a nbd $D(\pi(p))$ of $\pi(q)$ in $F$,
such that $\pi$ restricts from $D(p) \rightarrow D(\pi(p)$ and this restriction is bijetive with continuous inverse.
\end{lema}
\emph{Proof.}  

Let $p'=(u_0, v_0, t_0) \in \mathcal{F}$ and $D(\sigma(p')$ be the nbd of $\sigma(p')$ in $\mathcal{C}_F$ on which 
$\sigma^{-1}: D(\sigma(p')) \rightarrow \mathcal{F}$ is continuous.
As in Section~\ref{lsiSec}, we make the simplifying assumption that $f_t \neq 0$ at $p'$ (recall that,
$\mathcal{F}$ is defined as the zero-set of $f$). By the implicit function theorem there exists a nbd $D(q')$ of $q'=(u_0, v_0)$ and a function
$g: D(q') \rightarrow \mathbb{R}$ such that  $\forall (u,v) \in D(q')$, $f(u,v,g(u,v))=0$. 

\hfill $\square$
}

\subsection{Local similarity across faces}
We begin by studying the variation of the unit normal across faces of $\mathcal{C}$. Let $C_t$ denote the
curve of contact of $\mathcal{C}$ at time $t$. Let $p \in C_t$ be such that $p$ is common to (only) $\mathcal{C}_{F_1}$ and
$\mathcal{C}_{F_2}$ where $F_1$ and $F_2$ are two distinct faces of $M$.

\begin{lema} The faces $F_1$ and $F_2$ are adjacent in $M$. Further, 
the normal to $\mathcal{C}_{F_1}$ at $p$ is identical to the normal to $\mathcal{C}_{F_2}$ at $p$.
\end{lema}
\emph{Proof.} Suppose $p$ corresponds to $q$. Clearly, $q$ is common to (only) $F_1$ and $F_2$. Thus $F_1$ and $F_2$ are adjacent in $M$.
Further, by $G^1$ continuity, the normal to $F_1$ at $q$ is identical to the normal to $F_2$ at $q$. By lemma~\ref{localnormallemma}, it is
clear that, the normals to $\mathcal{C}_{F_1}$, and to $\mathcal{C}_{F_2}$, at $p$ are identical.
\hfill $\square$

Thus, the adjacencies on $\mathcal{C}$ are the `same' as the adjacencies on $M$. See figure~\ref{sweepExample}, which effectively
illustrates this through colours.  \eat{Clearly, the map $\pi$ restricts naturally from $\mathcal{C}_F \rightarrow F$.}
Further, the adjacent entities of $\mathcal{C}$
meet smoothly across common lower-dimensional entities. In other words, $\mathcal{C}$ is also of class $G^1$. 
Recall that the overall envelope may be obtained from the contact-set $\mathcal{C}$
by simply capping the appropriate parts of the solid $M$ at the initial and final position.
Therefore, the topological structure of the contact-set and hence, that of the envelope, mimics that of the solid.

The following theorem summarizes the analysis so far.
\begin{thrm} The map $\pi$ from the contact-set/envelope to the solid is an adjacency-respecting local homeomorphism 
onto its image. 
\end{thrm}

\subsection{Curvature of $\mathcal{C}$} \label{curvatureSubSec}
In the special case when the trajectory $h$ consists only of translations, i.e. $A(t)=I$ $\forall t$, the Gaussian curvature of $\mathcal{C}$ can be 
expressed in terms of the Gaussian curvature of $S$ and the curvature of $h$.  Since $A(t) = I$ $\forall t$, $\sigma_u = S_u$ and $\sigma_v = S_v$. 
Also, $V_u = V_v = 0$.  Consider a point $p = (u_0, v_0, t_0) \in \mathcal{F}$.  By definition of $\mathcal{F}$, we have that $f(u_0, v_0, t_0)=0$.  
Given that $\nabla f|_p \neq 0$ suppose without loss of generality that $f_t|_p \neq 0$.  Then by the implicit function theorem there exists a neighbourhood 
$\mathcal{N}$ of $q=(u_0, v_0)$ such that  $\forall (u,v) \in \mathcal{N}$, $f(u,v,t(u,v))=0$.  Hence, $t_u = -\frac{f_u}{f_t}$ and $t_v = -\frac{f_v}{f_t}$ 
and we get a local parameterization of $\mathcal{C}$ in $\mathcal{N}$ by $\psi(u,v) = \sigma(u,v,t(u,v))$.  $\mathcal{T}_{\mathcal{C}}(p)$ is spanned by 
$\psi_u = \sigma_u + \sigma_t t_u$ and $\psi_v = \sigma_v + \sigma_t t_v$.  Since $p \in \mathcal{F}$ by lemma~\ref{envlLem}, $\sigma_t$ is in the 
space spanned by $\sigma_u$ and $\sigma_v$.  Let $\sigma_t = l \sigma_u + m \sigma_v$.  Hence we express basis for $\mathcal{T}_{\mathcal{C}}(p)$ 
in terms of basis of $\mathcal{T}_{S}(q)$ as follows
\begin{align}  \label{Meq}
\begin{bmatrix}
\psi_u &  \psi_v 
\end{bmatrix}
= 
\begin{bmatrix}
\sigma_u &  \sigma_v 
\end{bmatrix}
\mathbf{M}
\end{align}
\noindent where $\mathbf{M} = \begin{bmatrix}   1+l t_u & l t_v \\  m t_u & 1+m t_v  \end{bmatrix}$.
The unit normal to $\mathcal{C}$ is given by $\hat{N}(u,v) = A(t(u,v))N(u,v) = N(u,v)$ where $N$ is the unit normal to $S_t$.  Hence, $\hat{N}_u = N_u$ and $\hat{N}_v = N_v$. Further, 
\begin{align} \label{Weq}
\begin{bmatrix}
N_u & N_v
\end{bmatrix}
=
\begin{bmatrix} 
\sigma_u & \sigma_v
\end{bmatrix}
\mathbf{W}
\end{align}
\noindent where $\mathbf{W}$ is the Weingarten matrix of $S$ at point $q$ whose determinant gives the Gaussian curvature of $S$ at $q$(see ~\cite{eleDiffGeo}).  From Eq.~\ref{Meq} and Eq.~\ref{Weq} we have
\begin{align}
\begin{bmatrix}
\hat{N}_u & \hat{N}_v
\end{bmatrix}
=
\begin{bmatrix}
\psi_u & \psi_v
\end{bmatrix}
\mathbf{M}^{-1}\mathbf{W}
\end{align}

So the Weingarten matrix of $\mathcal{C}$ with respect to parameterization $\psi$ is given by $\mathbf{M}^{-1}\mathbf{W}$ and the Gaussian curvature is given by $det(\mathbf{M}^{-1}\mathbf{W})$.  From Eq.~\ref{envlCondEq} and the fact that $N_t = V_u = V_v = 0$ we note that $f = \left < N, V \right >$, $f_u = \left < N_u, V \right>$, $f_v = \left < N_v, V \right>$ and $f_t = \left <N, V_t \right>$.
So,
\begin{align*}
det(\mathbf{M}^{-1}) &= \frac{1}{1+l t_u + m t_v} = \frac{f_t}{f_t - l f_u - m f_v} \\
			&= \frac{\left< N, V_t \right>}{ \left<N, V_t \right> - \left< l N_u + m N_v, V \right>} \\
			&= \frac{\left< N, V_t \right>}{ \left<N, V_t \right> - \left < \mathbf{W}(V), V \right >}
\end{align*} 
The Gaussian curvature of $\mathcal{C}$ is computed as
\begin{align}
det(\mathbf{M}^{-1}\mathbf{W}) &= \frac{\left< N, V_t \right> }{ \left<N, V_t \right> - \left < \mathbf{W}(V), V \right >} det(\mathbf{W})
\end{align}
\noindent where, $\left < N, V_t \right>$ is the curvature of the trajectory scaled by $\| V_t \|$, $\left < \mathbf{W}(V), V \right >$ is the normal curvature of $S$ at $p$ in direction $V$ scaled by $\|V\|^2$ and $det(\mathbf{W})$ is the Gaussian curvature of $S$ at $p$.

\section{Envelope computation} \label{computationSec}

In this section we describe the construction of the envelope $\mathcal{E}$ assuming that it is free from self-intersections and hence regular.  
We obtain a procedural parametrization of $\mathcal{E}$.  The procedural paradigm is an abstract way of defining curves and surfaces.  It relies on the fact that from 
the user's point of view, a parametric surface(curve) in $\mathbb{R}^3$ is a map from $\mathbb{R}^2 (\mathbb{R})$ to $\mathbb{R}^3$ and 
hence is merely a set of programs which allow the user to query the key attributes of the surface(curve), e.g. its domain and to evaluate the 
surface(curve) and its derivatives at the given parameter value.  The procedural approach to defining geometry is especially useful when closed 
form formulae are not available for the parametrization map and one must resort to iterative numerical methods.  We use the Newton-Raphson(NR) 
method for this purpose. As an example, the parametrization of the intersection curve of two surfaces is computed procedurally in \cite{procedural}.  As we will see, 
this approach has the advantage of being computationally efficient as well as accurate.  For a detailed discussion on the procedural framework, see~\cite{sohoni}.
  
The computational framework is as follows.  For the input parametric surface $S$ and trajectory $h$, an approximate envelope is first computed, 
which we will refer to as the seed surface.  Now, when the user wishes to evaluate the actual envelope or its derivative at some parameter value,  
a NR method will be started with seed obtained from the seed surface.  The NR method will converge, upto the required tolerance, to the required 
point on the envelope, or to its derivative, as required.  Here, the precision of the evaluation is only restricted by the finite precision of the computer
 and hence is accurate.  It has the advantage that if a tighter degree of tolerance is required while evaluation of the surface or its derivative, the seed 
surface does not need to be recomputed.  Thus, for the procedural definition of the envelope we need the following:
\begin{enumerate}
\item a NR formulation for computing points on $\mathcal{E}$ and its derivatives, which we describe in subsection~\ref{NRFormSubSec}
\item Seed surface for seeding the NR procedure, which we describe in subsection~\ref{seedSubSec}
\end{enumerate}

Recall that by the non-degeneracy assumption, $\mathcal{E}$ is the union of $C_t, \forall t$.  This suggests a natural parametrization of $\mathcal{E}$ in 
which one of the surface parameters is time $t$.  We will call the other parameter $p$ and denote the seed surface by $\gamma$ which is a map 
from the parameter space of $\mathcal{E}$ to the parameter space of $\mathcal{\sigma}$, i.e. $\gamma(p,t) = (\bar{u}(p,t), \bar{v}(p,t), t)$ and while 
the point $\sigma(\gamma(p,t))$ may not belong to $\mathcal{E}$, it is close to $\mathcal{E}$.  In other words, $\gamma(p,t)$ is close to $\mathcal{F}$.  
We call the image of the seed surface through the sweep map $\sigma$ as the approximate envelope and denote it by $\bar{\mathcal{E}}$, 
i.e. $\bar{\mathcal{E}}(p,t) = \sigma(\gamma(p,t))$.  We make the following assumption about $\bar{\mathcal{E}}$.
\begin{assum} \label{oneOneAssum}
At every point on the \emph{iso-t} curve of $\bar{\mathcal{E}}$, the normal plane to the \emph{iso-t} curve intersects the \emph{iso-t} curve of $\mathcal{E}$ in exactly one point.
\end{assum}
Note that this is not a very strong assumption and holds true in practice even with rather sparse sampling of points for the seed surface.  We now describe the Newton-Raphson formulation for evaluating points on $\mathcal{E}$ and its derivatives at a given parameter value.

\subsection{NR formulation for faces of $\mathcal{E}$} \label{NRFormSubSec}

Recall that the points on $\mathcal{E}$ were characterized by the tangency condition given in Eq.~\ref{envlCondEq}.  Introducing the parameters $(p,t)$ of $\mathcal{E}$, we rewrite Eq.~\ref{envlCondEq} $\forall (p_0, t_0)$:
\begin{align} \label{envlCondParEq}
\nonumber f(u(p_0,t_0), v(p_0,t_0), t_0 )  &= \left < \hat{N}(u(p_0,t_0), v(p_0,t_0), t_0), \right . \\
			&\left . V(u(p_0,t_0), v(p_0,t_0), t_0) \right > = 0
\end{align}
So, given $(p_0,t_0)$, we have one equation in two unknowns, viz. $u(p_0,t_0)$ and $v(p_0, t_0)$. $\mathcal{E}(p_0,t_0)$ is defined as the 
intersection of the plane normal to the iso-$t$(for $t=t_0$) curve of $\bar{\mathcal{E}}$ at $\bar{\mathcal{E}}(p_0,t_0)$ with the iso-$t$(for $t=t_0$) 
curve of $\mathcal{E}$ which is nothing but $C_{t_0}$. Recall that $C_{t_0}$ is given by $\sigma(u(p, t_0), v(p, t_0), t_0)$ where $u, v, t$ obey Eq.~\ref{envlCondParEq}.  
Henceforth, we will suppress the notation that $u,v, \bar{u}$ and $\bar{v}$ are functions of $p$ and $t$.  Also, all the evaluations will be 
understood to be done at parameter values $(p_0,t_0)$.  The tangent to iso-$t$ curve of $\bar{\mathcal{E}}$ at $(p_0,t_0)$  is given by 
\begin{align}
\frac{\partial \bar{\mathcal{E}}}{\partial p} =\frac{\partial \sigma}{\partial u} \frac{\partial \bar{u}}{\partial p} + \frac{\partial \sigma}{\partial v} \frac{\partial \bar{v}}{\partial p}
\end{align}
Hence, $\mathcal{E}(p_0,t_0)$ is the solution of simultaneous system of equations~\ref{envlCondParEq} and~\ref{planeOrthoEq}
\begin{align} \label{planeOrthoEq}
\left < \sigma(u, v ,t_0) - \sigma(\bar{u}, \bar{v}, t_0)   ,   \frac{\partial \bar{\mathcal{E}}}{\partial p} \right > = 0 
\end{align}
Eq.~\ref{envlCondParEq} and Eq.~\ref{planeOrthoEq} give us a system of two equations in two unknowns, $u$ and $v$ and hence can be put into NR 
framework by computing their first order derivatives w.r.t $u$ and $v$.  For any given parameter value $(p_0,t_0)$, we seed the NR method with the 
point $(\bar{u}(p_0,t_0), \bar{v}(p_0,t_0))$ and solve  Eq.~\ref{envlCondParEq} and Eq.~\ref{planeOrthoEq} for $(u(p_0,t_0), v(p_0,t_0))$ and compute $\mathcal{E}(p_0,t_0)$.

Having computed $\mathcal{E}(p,t)$ we now compute first order derivatives of $\mathcal{E}$ assuming that they exist.  In order to compute $\frac{\partial \mathcal{E}}{\partial p}$, we differentiate Eq.~\ref{envlCondParEq} and Eq.~\ref{planeOrthoEq} w.r.t. $p$ to obtain
\begin{align} 
&\left < \frac{\partial \hat{N}}{\partial u} \frac{\partial u}{\partial p} + \frac{\partial \hat{N}}{\partial v} \frac{\partial v}{\partial p},  V \right > + \left< \hat{N} , \frac{\partial V}{\partial u} \frac{\partial u}{\partial p} + \frac{\partial V}{\partial v} \frac{\partial v}{\partial p} \right >=0  \label{derPEq1}  \\
\nonumber &\left< \frac{\partial \sigma}{\partial u} \frac{\partial u}{\partial p} +  \frac{\partial \sigma}{\partial v} \frac{\partial v}{\partial p} - \frac{\partial \sigma}{\partial u} \frac{\partial \bar{u}}{\partial p} +  \frac{\partial \sigma}{\partial v} \frac{\partial \bar{v}}{\partial p},   \frac{\partial \bar{\mathcal{E}}}{\partial p}\right> \\
&+ \left < \sigma(u, v ,t_0) - \sigma(\bar{u}, \bar{v}, t_0) , \frac{\partial ^2 \bar{\mathcal{E}}}{\partial p^2} \right >= 0   \label{derPEq2}
\end{align}
Eq.~\ref{derPEq1} and Eq.~\ref{derPEq2} give a system of two equations in two unknowns, viz., $\frac{\partial u}{\partial p}$ and 
$\frac{\partial v}{\partial p}$ and can be put into NR framework by computing first order derivatives w.r.t. $\frac{\partial u}{\partial p}$ 
and $\frac{\partial v}{\partial p}$.  Note that Eq.~\ref{derPEq1} and Eq.~\ref{derPEq2} also involve $u$ and $v$ whose computation we have already described.
After computing $\frac{\partial u}{\partial p}$ and $\frac{\partial v}{\partial p}$, $\frac{\partial \mathcal{E}}{\partial p}$ can be computed as 
$\frac{\partial \sigma}{\partial u} \frac{\partial {u}}{\partial p} + \frac{\partial \sigma}{\partial v} \frac{\partial {v}}{\partial p}$.  $\frac{\partial \mathcal{E}}{\partial t}$ 
can similarly be computed by differentiating Eq.~\ref{envlCondParEq} and Eq.~\ref{planeOrthoEq} w.r.t. $t$.

\eat{
Similarly, in order to compute $\frac{\partial \mathcal{E}}{\partial t}$, we differentiate Eq.~\ref{envlCondParEq} and Eq.~\ref{planeOrthoEq} w.r.t. $t$ to obtain
\begin{align} 
&\left < \frac{\partial \hat{N}}{\partial u} \frac{\partial u}{\partial t} + \frac{\partial \hat{N}}{\partial v} \frac{\partial v}{\partial t},  V \right > + \left< \hat{N} , \frac{\partial V}{\partial u} \frac{\partial u}{\partial t} + \frac{\partial V}{\partial v} \frac{\partial v}{\partial t} \right >=0  \label{derTEq1}  \\
\nonumber &\left< \frac{\partial \sigma}{\partial u} \frac{\partial u}{\partial t} +  \frac{\partial \sigma}{\partial v} \frac{\partial v}{\partial t} - \frac{\partial \sigma}{\partial u} \frac{\partial \bar{u}}{\partial t} +  \frac{\partial \sigma}{\partial v} \frac{\partial \bar{v}}{\partial t},   \frac{\partial \bar{\mathcal{E}}}{\partial p}\right> \\
&+ \left < \sigma(u, v ,t_0) - \sigma(\bar{u}, \bar{v}, t_0) , \frac{\partial ^2 \bar{\mathcal{E}}}{\partial p \partial t} \right >= 0   \label{derTEq2}
\end{align}
Eq.~\ref{derTEq1} and Eq.~\ref{derTEq2} give a system of two equations in two unknowns, viz., $\frac{\partial u}{\partial t}$ and $\frac{\partial v}{\partial t}$ and can be put into NR framework by computing first order derivatives w.r.t. $\frac{\partial u}{\partial t}$ and $\frac{\partial v}{\partial t}$. 
}
 Higher order derivatives can be computed in a similar manner.

\subsection{Computation of seed surface} \label{seedSubSec}

The seed surface is constructed by sampling a few points on the envelope and fitting a tensor product B-spline surface through these points.  For this, we first sample a few time instants, say, $T =\{t_1, t_2, \ldots, t_n \}$ from the time interval of the sweep.  For each $t_i \in T$,  we sample a few points on the curve-of-contact $C_{t_i}$.  For this, we begin with one point $p$ on $C_{t_i}$ and compute the tangent to $C_{t_i}$ at $p$, call it $\mathcal{T}_{C_{i}}(p)$.  $p+\mathcal{T}_{C_{i}}(p)$ is used as a seed in Newton-Raphson method to obtain the next point on $C_{t_i}$ and this process is repeated.

While we do not know of any structured way of choosing the number of sampled points, in practice even a small number of points suffice to ensure that the Assumption~\ref{oneOneAssum} is valid.

\subsection{NR formulation for edges and vertices of $\mathcal{E}$}
An NR formulation for edges and vertices of $\mathcal{E}$ can be obtained in a manner similar to that for faces of $\mathcal{E}$ which we described in subsection~\ref{NRFormSubSec}.  In order to obtain a procedural parametrization for edges of $\mathcal{E}$, again seed curves need to be computed.

\section{Discussion} \label{conclusion}
In this work, we have proposed a novel computationally efficient test for detecting
anomalies on 
the envelope. This has been achieved through a delicate mathematical analysis of an
`invariant'.
We have provided a rich procedural framework for computing the Brep of the envelope
along with its
accurate parametrization. Another contribution is a natural correspondence between the 
geometric/topological entities of the Brep envelope and that of the Brep solid. This
framework has
been implemented using the ACIS kernel~\cite{acis} and has been used to produce the
running examples of
this paper.

Ongoing work includes, for example, extending the proposed procedural framework to
handle 
(i) deeper topological information of the Brep envelope, 
(ii) swept edges (an edge on the solid sweeping a face on the envelope) and so on 
(iii) faces meeting with $G^0$ continuity (i.e. sharp edges).
We also to plan to extend the detection of anomalies to above settings and further,
trim 
the appropriate part to obtain the final envelope. 

Another exciting future direction would be to analyse sweeps in which some numerical
invariant associated with a
a curve of contact, varies over time. For example, one may imagine sweeping a torus
along a trajectory where 
the number of components of the curve of contact changes over time. In such a case,
one would like to efficiently 
compute deeper topological invariants, say genus, of the envelope.
Our mathematical analysis coupled with a Morse-theoretic analysis appears promising.



\end{document}